\documentclass[11pt]{article}
\baselineskip
\pdfoutput=1
\usepackage{amsmath}
\usepackage{amssymb}
\usepackage{graphicx}
\usepackage{color}
\usepackage{cite}
\usepackage{collref}
\usepackage{tabls}
\usepackage{hyperref}
\DeclareMathOperator{\Tr}{Tr}
\DeclareMathOperator{\real}{Re}
\DeclareMathOperator{\diag}{diag}

\newcommand{\SU}{\mathrm{SU}}
\newcommand{\dd}{{\rm{d}}}
\newcommand{\alphas}{\alpha_{\mbox{\rm s}}}
\newcommand{\alphasMSbar}{\alpha_{\overline{\rm MS}}}
\newcommand{\tin}{t_{\mbox{\tiny{in}}}}
\newcommand{\tfin}{t_{\mbox{\tiny{fin}}}}
\newcommand{\Scl}{S_{\mbox{\tiny{cl}}}}
\newcommand{\ntraj}{n_{\mbox{\tiny{traj}}}}
\newcommand{\nqq}{n_{\mbox{\tiny{qq}}}}

\begin{document}
\begin{titlepage}
\renewcommand\thefootnote{\mbox{$\fnsymbol{footnote}$}}
\begin{center}
{\Large\bf Strong coupling from non-equilibrium Monte~Carlo simulations}
\end{center}
\vskip1.3cm
\centerline{Olmo~Francesconi,$^{a,b}$\footnote{\href{mailto:o.francesconi.961603@swansea.ac.uk}{{\tt o.francesconi.961603@swansea.ac.uk}}} Marco~Panero,$^{c,d}$\footnote{\href{mailto:marco.panero@unito.it}{{\tt marco.panero@unito.it}}} and David~Preti$^{d}$\footnote{\href{mailto:david.preti@to.infn.it}{{\tt david.preti@to.infn.it}}}}
\vskip1.5cm
\centerline{\sl $^a$Physics Department, College of Science, Swansea University (Singleton Campus)}
\centerline{\sl Swansea SA2 8PP, United Kingdom}
\vskip0.5cm
\centerline{\sl $^b$Universit\'e Grenoble Alpes, CNRS, LPMMC}
\centerline{\sl 38000 Grenoble, France}
\vskip0.5cm
\centerline{\sl $^c$Department of Physics, University of Turin and $^d$INFN, Turin}
\centerline{\sl Via Pietro Giuria 1, I-10125 Turin, Italy}
\vskip1.0cm

\setcounter{footnote}{0}
\renewcommand\thefootnote{\mbox{\arabic{footnote}}}
\begin{abstract}
\noindent We compute the running coupling of non-Abelian gauge theories in the Schr\"odinger-functional scheme, by means of non-equilibrium Monte~Carlo simulations on the lattice.
\end{abstract}

\end{titlepage}

\section{Introduction}
\label{sec:intro}

During the past few years there has been significant progress towards the understanding of quantum systems out of equilibrium and of the interplay between quantum and thermodynamics effects. Research combining theoretical tools from statistical mechanics, conformal field theory, the theory of integrable systems, and quantum information has led to a deeper comprehension of the connection between entanglement entropy and thermodynamic entropy in stationary states~\cite{Deutsch:2013moo, Kaufman:2016qtt}, as well as a clarification of the mechanism determining the time evolution of entanglement in many-body quantum systems out of equilibrium~\cite{Alba:2017lvc}.

At the same time, powerful fluctuation theorems were discovered and extensively studied in classical statistical mechanics (see refs.~\cite{Ritort:2004wf, MariniBettoloMarconi:2008fd, Esposito:2009zz} for reviews), that encode analytical relations among quantities characterizing systems driven out of thermodynamic equilibrium. These include the transient fluctuation theorem describing the probability of violations of the second law of thermodynamics in non-equilibrium steady states~\cite{Evans:1993po, Evans:1993em, Gallavotti:1994de, Gallavotti:1995de} and Jarzynski's identity, relating the free-energy difference between two equilibrium states of a system to the exponential average of the work done on the system to drive it out of equilibrium~\cite{Jarzynski:1996oqb, Jarzynski:1997ef}.

In the present work, we show how the latter theorem can be applied to study the renormalized coupling in non-Abelian, non supersymmetric gauge theory. This quantity is of major relevance in elementary particle theory: in particular, the gauge coupling $g$ of quantum chromodynamics (QCD) is one of the fundamental parameters in the Standard Model and plays a central r\^ole in theoretical predictions relevant for the physics probed in high-energy experiments\footnote{The most striking feature of the physical QCD coupling is its dependence on the momentum scale $\mu$: the dimensionless parameter $\alphas = g^2/(4\pi)$ is a \emph{decreasing} function of $\mu$~\cite{Gross:1973id, Politzer:1973fx}, so that QCD becomes a free theory at asymptotically high energies, while its behavior at low energies is non-perturbative. Note that the logarithmic running of the strong coupling is such that QCD remains a self-consistent theory for arbitrarily high energies~\cite{Brambilla:2014jmp}: a behavior remarkably different from other theories, like quantum electrodynamics, which break down at some high, but finite, energy scale.} like those at the CERN~LHC~\cite{Actis:2010gg, Dittmaier:2011ti, Heinemeyer:2013tqa, Dawson:2013bba, Adams:2013qkq}.

Specifically, we study the scale dependence of the gauge coupling in a non-equilibrium generalization of a Monte~Carlo calculation in the lattice regularization~\cite{Wilson:1974sk} by defining the theory in a four-dimensional box of finite linear extent $L$, with boundary conditions enforcing a non-trivial minimal-Euclidean-action configuration, and monitoring the response of the system under a sequence of quantum quenches (in Monte~Carlo time) that deform the boundary conditions driving the system out of equilibrium.

The terminology that we are using here is inspired by condensed-matter theory, where quantum quenches are a convenient tool to study systems driven out of equilibrium. A quantum quench is defined as a sudden change of the Hamiltonian of a quantum many-body system~\cite{Calabrese:2005in, Calabrese:2006rx}; before the quench, the system is in the ground state of the initial Hamiltonian, and the dynamical evolution after the quench is unitary. The system response to a quantum quench allows one to study many interesting aspects of its dynamics, including those related to localization~\cite{Serbyn:2014qq}, thermalization~\cite{Rigol:2007ta}, the interplay between integrable and non-integrable dynamics~\cite{Mussardo:2013cea}, and entanglement entropy~\cite{Calabrese:2007rg}. What makes quantum quenches particularly interesting is that, beside their theoretical interest, they can also be realized experimentally in certain condensed-matter systems~\cite{Mitra:2018wkj}. 

Here, we apply an analogous idea in the context of numerical simulations of a non-Abelian gauge theory regularized on the lattice; instead of changing the (bulk) Hamiltonian, we change the Dirichlet boundary conditions that are imposed on the fields along one of the four Euclidean directions of the system, and, instead of studying the real-time evolution induced by this change, we study its evolution in Monte~Carlo time. For technical reasons (related to the algorithm efficiency, to be discussed below), we apply not only one, but a sequence of such quenches. We then extract the physical gauge coupling at the length scale defined by the system size from the response of the system (as encoded in its quantum effective action) under the sequence of quenches that drives it out of equilibrium. The formalism rests directly on the definition of the coupling in the Schr\"odinger-functional scheme~\cite{Symanzik:1981wd, Luscher:1992an}, whereby the inverse squared physical coupling at distance $L$ is given (up to a normalization factor) by the derivative of the quantum effective action with respect to the parameter, to be denoted as $\eta$, specifying the Dirichlet boundary conditions. In other words, the Schr\"odinger-functional coupling is defined as a coefficient that encodes the response of the theory to a variation of the field enforced by the boundary conditions. 
It is important to note that, while the quantum effective action of the theory cannot be directly accessed in Monte~Carlo calculations (neither by conventional algorithms, nor by the one we discuss in this work), its derivative with respect to $\eta$ is a ``measurable'' quantity in a simulation, and, as we will discuss in detail below, our algorithm estimates numerically precisely this quantity, which is computed by means of Jarzynski's equality. For the sake of simplicity, the calculation is carried out in the pure-glue sector, for $\SU(2)$ and $\SU(3)$ gauge groups, and we show that the results obtained are fully compatible with previous calculations in the conventional (equilibrium) setting~\cite{Luscher:1992zx, Luscher:1993gh}. For the $\SU(2)$ theory, another relevant work was reported in ref.~\cite{deDivitiis:1994yz}, which studied the approach to the continuum with and without boundary improvement terms. We remark that the generalization to include dynamical matter fields and/or to other non-Abelian gauge groups is straightforward.

\section{Numerical implementation}
\label{sec:numerical_implementation}

Jarzynski's theorem~\cite{Jarzynski:1996oqb, Jarzynski:1997ef} states that when a thermodynamic system, initially in thermal equilibrium at temperature $T$, is driven out of equilibrium by a time-dependent variation protocol for the parameters $\lambda$ (such as couplings, etc.) of its Hamiltonian $H$ during a finite time interval $[\tin, \tfin]$, the exponential average of the work $W$ done on the system in units of the temperature is equal to the ratio of the partition functions (denoted by $Z$) for \emph{equilibrium} states of the system with parameters $\lambda(\tfin)$ and $\lambda(\tin)$:
\begin{equation}
\label{Jarzynski_theorem}
\left\langle \exp \left( -W/T \right) \right\rangle = \frac{Z_{\lambda(\tfin)}}{Z_{\lambda(\tin)}}.
\end{equation}
The quantity on the left-hand side of eq.~(\ref{Jarzynski_theorem}) is a statistical average over all possible evolutions of the system, when its parameters are modified according to a protocol $\lambda(t)$, which is \emph{fixed} and \emph{arbitrary}.

To clarify the meaning of the average appearing on the left-hand side of eq.~(\ref{Jarzynski_theorem}), and to describe how our algorithm works, it is convenient to review the proof of eq.~(\ref{Jarzynski_theorem}) in a setup that is relevant for our calculation. Note that the time $t$ that appears in Jarzynski's theorem can be either real time or Monte~Carlo time; in the following, we identify $t$ with Monte~Carlo time. We remark that the identity encoded in eq.~(\ref{Jarzynski_theorem}) is valid under general conditions; in particular, it holds when the starting configurations in the evolution of the system are at equilibrium and the dynamics of the system preserves the equilibrium distribution if the parameters $\lambda$ are fixed. To prove eq.~(\ref{Jarzynski_theorem}) for a statistical system undergoing Monte~Carlo evolution which satisfies the stronger detailed-balance condition (like the lattice version of the gauge theories considered in this study), let $\phi$ denote the degrees of freedom of the system, let
\begin{equation}
\label{normalized_equilibrium_distribution}
\pi_\lambda(\phi)=\exp[-H_\lambda(\phi)/T]/Z_\lambda
\end{equation}
be the normalized equilibrium distribution corresponding to a given, fixed value of $\lambda$ (and $Z_\lambda$ the corresponding partition function), and let $\mathcal{P}_\lambda(\phi \to \phi^\prime)$ be the normalized transition probability density from a given configuration $\phi$ to another configuration $\phi^\prime$ at fixed $\lambda$. The detailed-balance condition reads
\begin{equation}
\label{detailed_balance}
\pi_\lambda(\phi) \mathcal{P}_\lambda(\phi \to \phi^\prime) = \pi_\lambda(\phi^\prime) \mathcal{P}_\lambda(\phi^\prime \to \phi).
\end{equation}
In general, the work done on the system when the parameters are varied according to the  $\lambda(t)$ protocol (and, as a consequence, the fields undergo a non-equilibrium evolution $\phi(t)$) is
\begin{equation}
\label{work}
W=\int_{\tin}^{\tfin} \frac{\partial H}{\partial \lambda} \dot{\lambda} \dd t,
\end{equation}
where $\dot{\lambda}$ is the derivative of $\lambda$ with respect to $t$. In the rest of this article, following the terminology of refs.~\cite{Jarzynski:1996oqb, Jarzynski:1997ef}, we call each non-equilibrium evolution of the fields in Monte~Carlo time $\phi(t)$ a non-equilibrium Monte~Carlo \emph{trajectory}; as stated above, during their evolution the fields do not remain in equilibrium, because the parameters $\lambda$ are varied as a function of Monte Carlo time. Note that, in this context, the concept of trajectory is distinct from the one of a trajectory of the hybrid Monte~Carlo algorithm~\cite{Callaway:1982eb, Callaway:1983ee, Polonyi:1983tm, Duane:1985hz, Duane:1986iw}, which is widely used in the lattice QCD literature (even though, as will be briefly mentioned in section~\ref{sec:conclusions}, the non-equilibrium algorithm discussed in this work could \emph{also} be implemented in a non-equilibrium version of the hybrid Monte~Carlo algorithm). Let us assume that each Monte~Carlo trajectory is made of $n$ steps, and that at each step along a trajectory, $\lambda$ is first updated, then the field configuration is let evolve according to the dynamics defined by the new value of $\lambda$. Let $i$ label Monte~Carlo time (with $i=0$ corresponding to the initial time $\tin$ and $i=n$ corresponding to the final time $\tfin$), so that $\phi_i$ denotes the field configuration at Monte~Carlo time labeled by $i$. The work done on the system when $\lambda$ is varied, say, from $\lambda_i$ to $\lambda_{i+1}$ (and \emph{before} the Monte~Carlo update of the field), is the difference $H_{\lambda_{i+1}}(\phi_i)-H_{\lambda_i}(\phi_i)$. Then, using eq.~(\ref{normalized_equilibrium_distribution}), the exponential of minus the work divided by $T$, which appears in the average on the left-hand side of eq.~(\ref{Jarzynski_theorem}), can be rewritten as
\begin{equation}
\exp \left( -W/T \right) = \prod_{i=0}^{n-1} \frac{Z_{\lambda_{i+1}} \pi_{\lambda_{i+1}}(\phi_i)}{Z_{\lambda_i} \pi_{\lambda_i}(\phi_i)}.
\end{equation}
On the left-hand side of eq.~(\ref{Jarzynski_theorem}), this quantity is averaged over all trajectories that $\phi$ can span in its Monte~Carlo evolution: this corresponds to integrating over all possible configurations at each step in the Monte~Carlo trajectory. The initial configurations are distributed according to the equilibrium distribution $\pi_{\lambda_0}$, while those at later Monte~Carlo times are further weighted by products of the transition probabilities $\mathcal{P}_{\lambda_{i+1}}(\phi_i \to \phi_{i+1})$. Thus, the left-hand side of eq.~(\ref{Jarzynski_theorem}) can be expressed as
\begin{equation}
\left\langle \exp \left( -W/T \right) \right\rangle = \int \dd \phi_0 \int \dd \phi_1 \int \dd \phi_2 \dots \int \dd \phi_n \pi_{\lambda_0}(\phi_0) \prod_{i=0}^{n-1} \frac{Z_{\lambda_{i+1}}\pi_{\lambda_{i+1}}(\phi_i)\mathcal{P}_{\lambda_{i+1}} (\phi_i \to \phi_{i+1})}{Z_{\lambda_i} \pi_{\lambda_i}(\phi_i)} .
\end{equation}
All partition functions appearing in the product cancel against each other, except for the one appearing in the denominator of the fraction in the first term, and the one in the numerator in the last factor, leaving $Z_{\lambda_n}/Z_{\lambda_0}$. Using eq.~(\ref{detailed_balance}), the previous equation can be rewritten as
\begin{equation}
\left\langle \exp \left( -W/T \right) \right\rangle = \frac{Z_{\lambda_n}}{Z_{\lambda_0}} \int \dd \phi_0 \int \dd \phi_1 \int \dd \phi_2 \dots \int \dd \phi_n  \pi_{\lambda_0}(\phi_0) \prod_{i=0}^{n-1} \frac{\pi_{\lambda_{i+1}}(\phi_{i+1})\mathcal{P}_{\lambda_{i+1}} (\phi_{i+1} \to \phi_i)}{\pi_{\lambda_i}(\phi_i)} ,
\end{equation}
so that also the $\pi$ distributions cancel against each other, except for $\pi_{\lambda_n}(\phi_n)/\pi_{\lambda_0}(\phi_0)$. Multiplying this quantity by the $\pi_{\lambda_0}(\phi_0)$ term in front of the product, one is left with
\begin{equation}
\left\langle \exp \left( -W/T \right) \right\rangle =  \frac{Z_{\lambda_n}}{Z_{\lambda_0}} \int \dd \phi_0 \int \dd \phi_1 \int \dd \phi_2 \dots \int \dd \phi_n \pi_{\lambda_n}(\phi_n) \prod_{i=0}^{n-1} \mathcal{P}_{\lambda_{i+1}} (\phi_{i+1} \to \phi_i).
\end{equation}
In the latter expression, $\phi_0$ appears only in the $\mathcal{P}_{\lambda_1} (\phi_1 \to \phi_0)$ factor, and the normalization of the transition probabilities implies $\int \dd \phi_0 \mathcal{P}_{\lambda_1} (\phi_1 \to \phi_0) = 1$. The same argument can then be repeated for the integrals over $\phi_1$, $\phi_2$, $\dots$, $\phi_{n-1}$. Finally, the last integral is $\int \dd \phi_n \pi_{\lambda_n}(\phi_n)$, which also equals one, because the equilibrium distributions are normalized, and one arrives at eq.~(\ref{Jarzynski_theorem}).

Let us now discuss how we implemented eq.~(\ref{Jarzynski_theorem}) in our algorithm for the numerical evaluation of the running coupling in the Schr\"odinger-functional scheme. Following refs.~\cite{Luscher:1992an, Luscher:1992zx, Luscher:1993gh}, we regularized the $\SU(N)$ Yang-Mills theory (with $N=2$ and $3$) on a hypercubic lattice of spacing $a$ and linear extent $L$ in each direction. The degrees of freedom of the theory (matrices $U$ in the defining representation of the gauge group) are associated with the oriented lattice links. Periodic boundary conditions are assumed along the three spatial directions, whereas fixed boundary conditions are imposed at the initial ($x_0=0$) and final ($x_0=L$) Euclidean time, where the spatial links are set to fixed, spatially uniform, Abelian matrices defined below, while no boundary conditions are imposed on the temporal links between sites on the boundaries and sites in the bulk of the lattice (and there are no positively oriented temporal links from the sites on the boundary at Euclidean time $x_0=L$, nor negatively oriented temporal links from the sites on the boundary at Euclidean time $x_0=0$). The dynamics is governed by the action $S = -(1/g_0^2) \sum_{p} w(p) \real \Tr U_p$~\cite{Wilson:1974sk}, where $g_0$ denotes the bare coupling, and $U_p$ is the path-ordered product of the matrices on the $a\times a$ square (``plaquette'') labeled by $p$. $w(p)=1$ in the bulk of the system, while it equals $1/2$ for spatial plaquettes on the three-dimensional slices at the Euclidean times $x_0=0$ and $x_0=L$, and it equals $c_t(g_0)$ for plaquettes parallel to the Euclidean-time direction and touching the boundaries. For consistency with the previous works we compare our results with, we set the ``improvement coefficient'' $c_t(g_0)$ to $1$ for $N=2$~\cite{Luscher:1992zx}, whereas $c_t(g_0)=1-0.089 g_0^2$ for $N=3$~\cite{Luscher:1993gh}. For later convenience, we also define $\beta=2N/g_0^2$.

The reformulation of eq.~(\ref{Jarzynski_theorem}) in the Euclidean quantum-field-theory setting relevant for our Monte~Carlo simulations is straightforward, with $W/T$ replaced by the total Euclidean-action variation $\Delta S$ during each non-equilibrium trajectory of the field configuration~\cite{Caselle:2016wsw, Caselle:2018kap}. Note that, since each non-equilibrium trajectory is decomposed into $n$ steps, so is the Euclidean-action variation $\Delta S$: more precisely,
\begin{equation}
\label{discretized_Euclidean-action_difference}
\Delta S = \sum_{i=0}^{n-1} \left[ S_{\lambda_{i+1}}(\phi_i)-S_{\lambda_{i}}(\phi_i)\right].
\end{equation}
In our calculations, $\lambda$ is identified with the angle $\eta$ that defines the field configurations for spatial link matrices at the boundaries, \emph{viz} $U=\exp(iaC_{x_0})$ with
\begin{equation}
\label{boundary_conditions_in_terms_of_eta}
C_0 = \frac{1}{L} \diag \left( -\eta , \eta \right), \qquad \qquad 
C_L = \frac{1}{L} \diag \left( \eta-\pi , \pi-\eta \right)
\end{equation}
for $N=2$ and 
\begin{eqnarray}
\label{boundary_conditions_in_terms_of_eta_and_nu}
&& C_0 = \frac{1}{L} \diag \left(
\eta - \frac{\pi}{3},
\eta \left( \nu - \frac{1}{2} \right),
-\eta \left( \nu + \frac{1}{2} \right) + \frac{\pi}{3}
\right), \nonumber \\
&& C_L = \frac{1}{L} \diag \left(
-\eta - \pi,
\eta \left( \nu + \frac{1}{2} \right) + \frac{\pi}{3},
-\eta \left( \nu - \frac{1}{2} \right) + \frac{2\pi}{3}
\right)
\end{eqnarray}
for $N=3$ (in the following, we set $\nu=0$). Classically, this induces a spatially uniform Abelian gauge field configuration with Euclidean action
\begin{equation}
\label{lattice_SU2_background_field_action}
\Scl = \frac{24 L^4}{g_0^2 a^4} \sin^2 \left[ \frac{a^2}{2L^2} (\pi-2\eta) \right]
\end{equation}
for $N=2$, and
\begin{equation}
\label{lattice_SU3_background_field_action}
\Scl = \frac{12 L^4}{g_0^2 a^4} \left\{ \sin^2 \left[ \frac{a^2}{L^2}\left( \eta + \frac{\pi}{3}\right) \right] + 2\sin^2 \left[ \frac{a^2}{2L^2}\left( \eta + \frac{\pi}{3}\right) \right] \right\}
\end{equation}
for $N=3$.

We define the evolution of $\lambda(t)$ as a sequence of $\nqq=n$ quantum quenches in Monte~Carlo time, in which $\eta$ is varied from an initial value $\eta(\tin)$ (equal to $\pi/4$ for $N=2$, or to $0$, for $N=3$) to a final value $\eta(\tfin)=\eta(\tin)+\Delta \eta$; for simplicity, the amplitude of these quenches is taken to be constant, $\Delta \eta/\nqq$. After each quench, the field configuration is changed by a Monte~Carlo step (which consists of one heat-bath~\cite{Creutz:1980zw, Kennedy:1985nu} and three to ten over-relaxation updates~\cite{Adler:1981sn, Brown:1987rra} on $\SU(2)$ subgroups~\cite{Cabibbo:1982zn} for all $U$ matrices): this is done \emph{without} allowing the field to thermalize, thus driving the configuration progressively out of equilibrium. We verified that a ``reverse'' implementation of this non-equilibrium evolution, from $\eta(\tfin)$ to $\eta(\tin)$, always yield consistent results: in view of the non-symmetric r\^oles of the initial and final states, this is a non-trivial check of the robustness of our calculation. We compute the $Z_{\lambda(\tfin)}/Z_{\lambda(\tin)}$ ratio using eq.~(\ref{Jarzynski_theorem}). Setting $\Gamma = -\ln Z$, the physical coupling at the length scale $L$ is then defined as the ratio between the derivative of $g_0^2\Scl$ with respect to $\eta$ and the derivative of $\Gamma$ with respect to $\eta$. In turn, the latter is given by the limit of the difference quotient $\Delta \Gamma / \Delta \eta$ for $\Delta \eta \to 0$, so that one obtains
\begin{equation}
\label{SU2_SF_coupling}
g^2(L) = - \lim_{\Delta \eta \to 0} \frac{24 \Delta \eta}{\Delta \Gamma} \left( \frac{L}{a} \right)^2 \sin \left[ \frac{\pi}{2} \left( \frac{a}{L} \right)^2 \right]
\end{equation}
for the $\SU(2)$ theory (having set $\eta=\pi/4$) and
\begin{equation}
\label{SU3_SF_coupling}
g^2(L) =  \lim_{\Delta \eta \to 0} \frac{12 \Delta \eta}{\Delta \Gamma} \left( \frac{L}{a} \right)^2 \left\{ \sin \left[ \frac{2\pi}{3} \left( \frac{a}{L} \right)^2 \right] + \sin \left[ \frac{\pi}{3} \left( \frac{a}{L} \right)^2 \right] \right\}
\end{equation}
in the $\SU(3)$ theory (with $\eta=\nu=0$).

It is worth remarking that the quality of the numerical estimate of the average on the left-hand side of eq.~(\ref{Jarzynski_theorem}) depends crucially on how far from equilibrium the field configurations are driven during the Monte~Carlo trajectories, and on the statistics of trajectories that are sampled. In a nutshell, the exponential average in eq.~(\ref{Jarzynski_theorem}) implies that arbitrarily large deviations from equilibrium would require prohibitively large statistics to probe the tail of the $\Delta S$ distribution. The present calculation, however, does not require to probe deep out-of-equilibrium dynamics, as the physical coupling is obtained in the limit of small $\Delta \eta$ (and, consequently, small deviations from equilibrium). The bounds on the number of trajectories required to achieve a given level of precision in experimental or numerical sampling of out-of-equilibrium distributions are mathematically well understood~\cite{Zuckerman:2002toa, Jarzynski:2006re, Pohorille:2010gp, YungerHalpern:2016no, Arrar:2019ota} and are always satisfied in our Monte~Carlo simulations.

Following the procedure outlined in refs.~\cite{Luscher:1992zx, Luscher:1993gh}, the evolution of the physical coupling as a function of the momentum scale $O(1/L)$ is then defined in an iterative way, in terms of the continuum-extrapolated step-scaling function $\sigma (s, g^2(L))=g^2(sL)$  that was introduced in ref.~\cite{Luscher:1991wu}. Note that $\sigma$ can be thought of as an integrated version of the $\beta$ function of the theory, as it describes the evolution of the coupling between the length scales $L$ and $sL$. We used $s=2$ and $s=3/2$.

\section{Results and analysis}
\label{sec:results_and_analysis}

\subsection{Results for the $\SU(2)$ theory}
\label{subsec:results_for_the_SU2_theory}

\begin{figure}[!htbp]
\begin{center}
\includegraphics*[width=\textwidth]{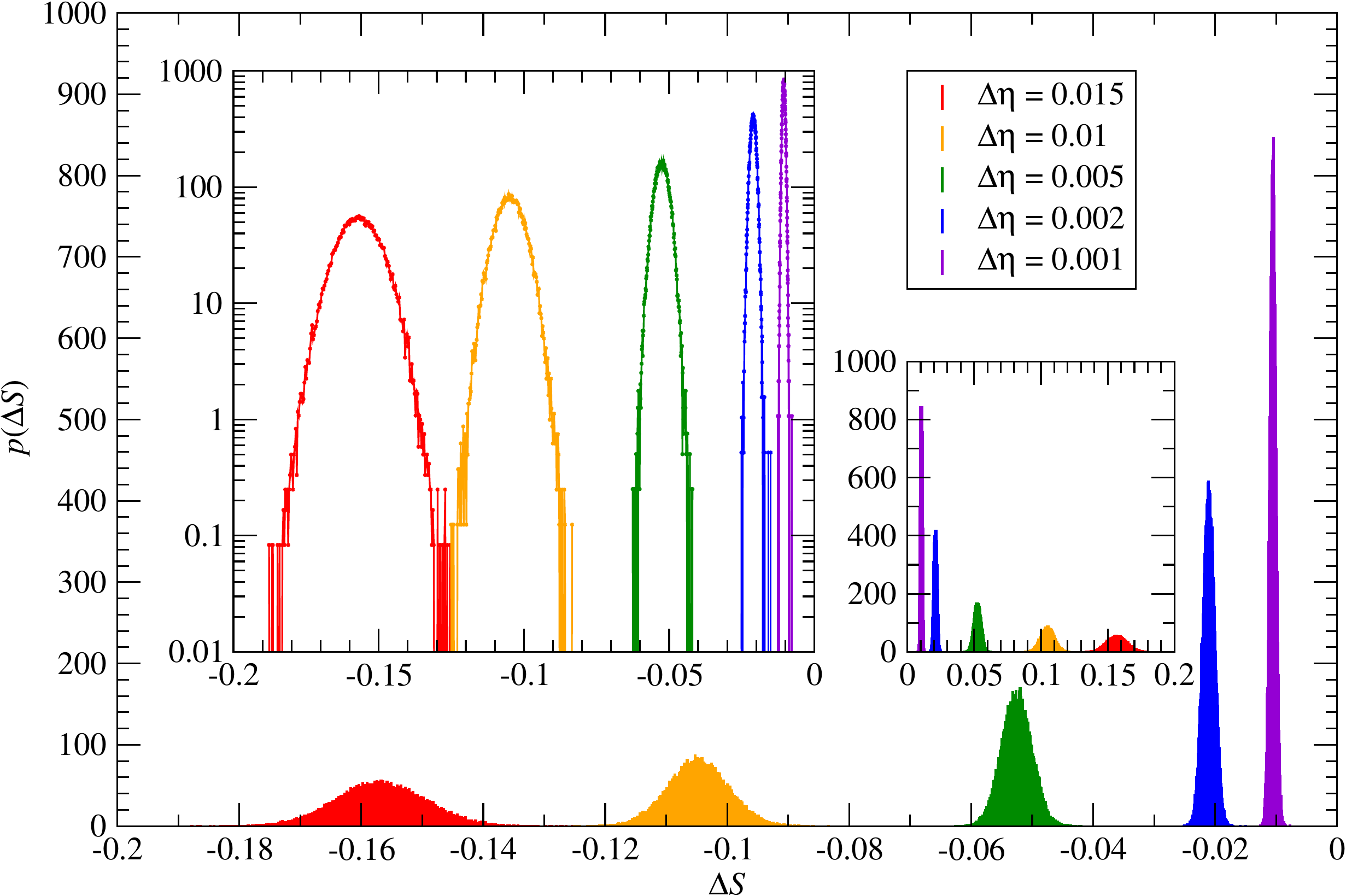}
\caption{\label{fig:plot_of_distribution_Ncol_2_nt_6_nx_5_ny_5_nz_5_beta_2.7124000000_nsteps_200_eta_0.7853981634} Distribution of the Euclidean action difference $\Delta S$ in non-equilibrium simulations of the $\SU(2)$ gauge theory in a hypercubic box of linear size $L=5a$ at $\beta=2.7124$, with the boundary fields specified in eq.~(\ref{boundary_conditions_in_terms_of_eta}). The histograms show the distribution of $\Delta S$ induced by a ``direct'' non-equilibrium transformation in which $\eta$ is varied from $\eta=\pi/4$ to $\eta=\pi/4+\Delta \eta$ through a sequence of $\nqq=200$ quenches, for different values of $\Delta \eta$. The larger inset shows the same distributions using a logarithmic scale for the vertical axis. In the smaller inset, the results from ``reverse'' transformations, from $\eta=\pi/4+\Delta \eta$ to $\eta=\pi/4$, are displayed.}
\end{center}
\end{figure}

We first discuss the $\SU(2)$ theory. The first step in the analysis of our numerical results consists in studying the distribution of Euclidean-action variations along the non-equilibrium trajectories. As an example, figure~\ref{fig:plot_of_distribution_Ncol_2_nt_6_nx_5_ny_5_nz_5_beta_2.7124000000_nsteps_200_eta_0.7853981634} shows the results obtained from simulations with $N=2$, $L=5a$ at $\beta=4/g_0^2=2.7124$, for different values of $\Delta \eta$ and $\nqq=200$ quenches. We note that the numerical results can be approximately modeled by Gau{\ss}ian distributions centered at $-0.156800(39)$ (for $\Delta \eta = 0.015$), at $-0.104812(26)$ (for $\Delta \eta = 0.01$), at $-0.052544(13)$ (for $\Delta \eta = 0.005$), at $-0.021054(5)$ (for $\Delta \eta = 0.002$), and at $-0.0105310(26)$ (for $\Delta \eta = 0.001$). The width of these distributions decreases to zero with $\Delta \eta$, as expected at fixed $\nqq$. As will be discussed in detail in subsection~\ref{subsec:computational_efficiency_analysis}, this is simply a consequence of the fact that, for very small values of $\Delta \eta/\nqq$, the field configurations remain close to equilibrium in every trajectory: for $\Delta \eta/\nqq=0$, the simulation would reduce to a conventional equilibrium Monte~Carlo. We also note that the distributions of Euclidean-action variations in reverse trajectories, from $\eta(\tin)=\pi/4+\Delta \eta$ to $\eta(\tfin)=\pi/4$, are approximately symmetric with respect to those observed in direct trajectories.

A more detailed analysis of the results displayed in figure~\ref{fig:plot_of_distribution_Ncol_2_nt_6_nx_5_ny_5_nz_5_beta_2.7124000000_nsteps_200_eta_0.7853981634}, providing information about the efficiency of our numerical algorithm, is presented in subsection~\ref{subsec:computational_efficiency_analysis}, together with a comparison with the computational costs of lattice calculations of the running coupling in the Schr\"odinger-functional scheme by means of conventional, equilibrium Monte~Carlo algorithm.

A summary of a larger sample of our data for the $\SU(2)$ gauge theory, for different values of $\Delta \eta$, and from direct and reverse implementations of our non-equilibrium Monte~Carlo simulations, is reported in table~\ref{tab:su2_results}. Here, the values of the bare coupling and system size are those in the last series of ref.~\cite[table 2]{Luscher:1992zx}, and the table shows the effective action difference $\Delta \Gamma=-\ln [Z_{\eta(\tfin)}/Z_{\eta(\tin)}]$. The table reveals clearly that direct and reverse non-equilibrium transformations yield compatible results, and that $\Delta \Gamma$ scales linearly with $\Delta \eta$.

\begin{figure}[!htbp]
\begin{center}
\includegraphics*[width=\textwidth]{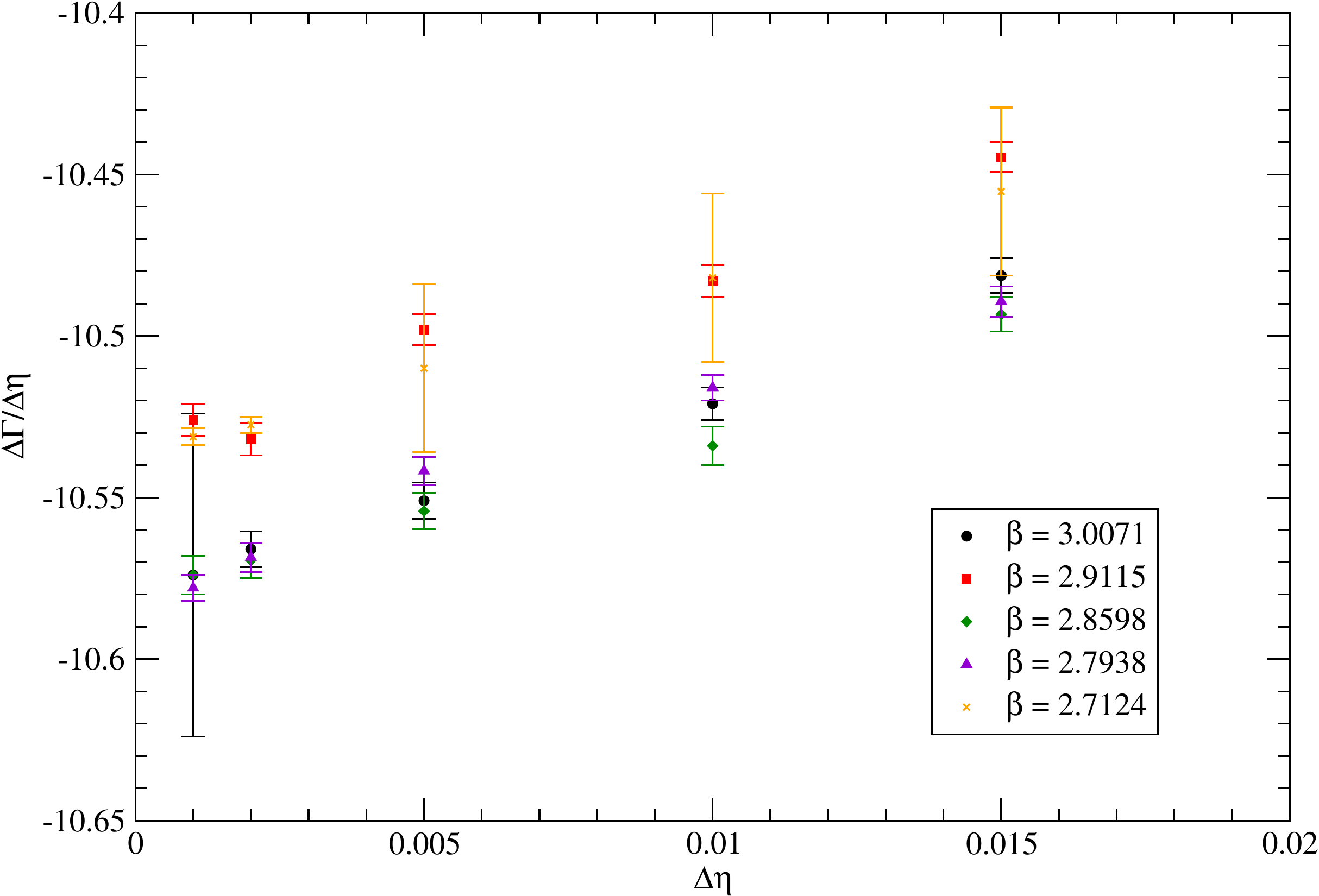}
\caption{\label{fig:SU2_Delta_Gamma_over_Delta_eta} The quotient ratio $\Delta \Gamma / \Delta \eta$, extracted from the simulation results listed in table~\ref{tab:su2_results}, as a function of $\Delta \eta$. The plot shows the results from ``direct'' non-equilibrium transformations at different $\beta$ values.}
\end{center}
\end{figure}

Figure~\ref{fig:SU2_Delta_Gamma_over_Delta_eta} shows the results for $\Delta \Gamma / \Delta \eta$ against $\Delta \eta$, as obtained from direct transformations at the different $\beta$ values reported in table~\ref{tab:su2_results}, corresponding to different values of the lattice spacing, for approximately constant physical linear size of the lattice: the plot reveals that the difference quotient remains essentially constant for all values of $\Delta \eta$. The figure shows the consistency of the results at different $\beta$ values, and a mild trend towards values of $\Delta \Gamma / \Delta \eta$ that are slightly more negative (i.e. larger in modulus) when $\Delta \eta$ is reduced towards zero, albeit only by an amount (with respect to the smallest $\Delta \eta$ considered here, i.e. $\Delta \eta=0.001$) that is comparable with our statistical uncertainties. This makes the evaluation of $g^2(L)$ from eq.~(\ref{SU2_SF_coupling}) robust and unambiguous. In view of these results, in order to reduce the computational cost of several statistically independent computations at different values of $\Delta \eta$, we then proceeded to the calculation of $g^2(L)$ from the results for $\Delta \Gamma$ obtained at $\Delta \eta=0.0001$ (a value ten times smaller than the smallest one used to produce the data sets reported in table~\ref{tab:su2_results} and shown in figure~\ref{fig:SU2_Delta_Gamma_over_Delta_eta}), increasing $\nqq$ to $1000$.

\begin{table}
\hspace{-16mm}
\begin{tabular}{|c|c|l|c|l|c|c|c|l|c|l|c|}
\hline
$\beta$ & $L/a$ & \multicolumn{1}{|c|}{$\Delta \eta$} & $\ntraj$ & \multicolumn{1}{|c|}{$\Delta \Gamma$} & type & \multicolumn{1}{||c|}{$\beta$} & $L/a$ & \multicolumn{1}{|c|}{$\Delta \eta$} & $\ntraj$ & \multicolumn{1}{|c|}{$\Delta \Gamma$} & type \\
\hline \hline
$2.7124$ & $5$ & $0.015$ & $36428$ & $-0.15683(39)$             & direct  & \multicolumn{1}{||c|}{$2.9115$} & $8$  & $0.015$ & $22620$ & $-0.15667(7)$             & direct  \\
         &     &         & $36448$ & $\phantom{+}0.15679(39)$   & reverse & \multicolumn{1}{||c|}{}         &      &         & $22627$ & $\phantom{+}0.15676(7)$   & reverse \\
         &     & $0.01$  & $36433$ & $-0.10482(26)$             & direct  & \multicolumn{1}{||c|}{}         &      & $0.01$  & $22625$ & $-0.10483(5)$             & direct  \\
         &     &         & $35209$ & $\phantom{+}0.10486(26)$   & reverse & \multicolumn{1}{||c|}{}         &      &         & $22626$ & $\phantom{+}0.10475(5)$   & reverse \\
         &     & $0.005$ & $36479$ & $-0.05255(13)$             & direct  & \multicolumn{1}{||c|}{}         &      & $0.005$ & $22519$ & $-0.052490(24)$           & direct  \\
         &     &         & $35885$ & $\phantom{+}0.05254(13)$   & reverse & \multicolumn{1}{||c|}{}         &      &         & $22634$ & $\phantom{+}0.052495(23)$ & reverse \\
         &     & $0.002$ & $36116$ & $-0.021055(5)$             & direct  & \multicolumn{1}{||c|}{}         &      & $0.002$ & $22628$ & $-0.021064(10)$           & direct  \\
         &     &         & $36325$ & $\phantom{+}0.021055(5)$   & reverse & \multicolumn{1}{||c|}{}         &      &         & $22628$ & $\phantom{+}0.021039(10)$ & reverse \\
         &     & $0.001$ & $36394$ & $-0.0105312(26)$           & direct  & \multicolumn{1}{||c|}{}         &      & $0.001$ & $22626$ & $-0.010526(5)$            & direct  \\
         &     &         & $36356$ & $\phantom{+}0.0105310(26)$ & reverse & \multicolumn{1}{||c|}{}         &      &         & $22638$ & $\phantom{+}0.010526(5)$  & reverse \\
\hline
$2.7938$ & $6$ & $0.015$ & $17534$ & $-0.15734(7)$              & direct  & \multicolumn{1}{||c|}{$3.0071$} & $10$ & $0.015$ & $25000$ & $-0.15722(8)$             & direct  \\
         &     &         & $17479$ & $\phantom{+}0.15740(7)$    & reverse & \multicolumn{1}{||c|}{}         &      &         & $25000$ & $\phantom{+}0.15742(8)$   & reverse \\
         &     & $0.01$  & $17531$ & $-0.10516(4)$              & direct  & \multicolumn{1}{||c|}{}         &      & $0.01$  & $25000$ & $-0.10521(5)$             & direct  \\
         &     &         & $17490$ & $\phantom{+}0.10520(4)$    & reverse & \multicolumn{1}{||c|}{}         &      &         & $25000$ & $\phantom{+}0.10530(6)$   & reverse \\
         &     & $0.005$ & $17555$ & $-0.052709(22)$            & direct  & \multicolumn{1}{||c|}{}         &      & $0.005$ & $25000$ & $-0.052755(28)$           & direct  \\
         &     &         & $17488$ & $\phantom{+}0.052748(22)$  & reverse & \multicolumn{1}{||c|}{}         &      &         & $25000$ & $\phantom{+}0.052761(28)$ & reverse \\
         &     & $0.002$ & $16998$ & $-0.021137(9)$             & direct  & \multicolumn{1}{||c|}{}         &      & $0.002$ & $25000$ & $-0.021132(11)$           & direct  \\
         &     &         & $17001$ & $\phantom{+}0.021123(9)$   & reverse & \multicolumn{1}{||c|}{}         &      &         & $25000$ & $\phantom{+}0.021153(11)$ & reverse \\
         &     & $0.001$ & $17008$ & $-0.010578(4)$             & direct  & \multicolumn{1}{||c|}{}         &      & $0.001$ & $25000$ & $-0.010574(5)$            & direct  \\
         &     &         & $16966$ & $\phantom{+}0.010569(4)$   & reverse & \multicolumn{1}{||c|}{}         &      &         & $25000$ & $\phantom{+}0.010569(5)$  & reverse \\
\hline
$2.8598$ & $7$ & $0.015$ & $13303$ & $-0.15740(8)$              & direct  & \multicolumn{6}{c}{} \\
         &     &         & $13312$ & $\phantom{+}0.15747(8)$    & reverse & \multicolumn{6}{c}{} \\
         &     & $0.01$  & $13302$ & $-0.10534(6)$              & direct  & \multicolumn{6}{c}{} \\
         &     &         & $13289$ & $\phantom{+}0.10520(6)$    & reverse & \multicolumn{6}{c}{} \\ 
         &     & $0.005$ & $13267$ & $-0.052771(28)$            & direct  & \multicolumn{6}{c}{} \\
         &     &         & $13299$ & $\phantom{+}0.052833(27)$  & reverse & \multicolumn{6}{c}{} \\
         &     & $0.002$ & $13292$ & $-0.021139(11)$            & direct  & \multicolumn{6}{c}{} \\ 
         &     &         & $13285$ & $\phantom{+}0.021146(11)$  & reverse & \multicolumn{6}{c}{} \\
         &     & $0.001$ & $13297$ & $-0.010574(6)$             & direct  & \multicolumn{6}{c}{} \\
         &     &         & $13293$ & $\phantom{+}0.010585(6)$   & reverse & \multicolumn{6}{c}{} \\
\cline{1-6}
\end{tabular}
\caption{Results for the effective-action variation $\Delta \Gamma$ in $\SU(2)$ gauge theory at five different combinations of $\beta$ and $L$ (corresponding to the last series reported in ref.~\cite[table 2]{Luscher:1992zx}) and for different values of $\Delta \eta$, in direct and in reverse non-equilibrium trajectories. These simulations were run with $\nqq=200$, with the number of trajectories denoted by $\ntraj$.}
\label{tab:su2_results}
\end{table}

\begin{table}
\begin{center}
\begin{tabular}{|c|c|c|r|l|r|l|}
\hline
$\beta$ & type & $L/a$ & $\ntraj(L)$ & \multicolumn{1}{|c|}{$g^2(L)$} & $\ntraj(2L)$ & \multicolumn{1}{|c|}{$g^2(2L)$} \\
\hline \hline
$3.4564$ & direct  &  $5$ & $3955408$ & $2.037933(14)$ & $358184$ & $2.43944(11)$ \\ 
         & reverse &      & $3956061$ & $2.037935(14)$ & $358304$ & $2.43936(11)$ \\ 
         & average &      &           & $2.037934(10)$ &          & $2.43940(8)$  \\ \hline
$3.5408$ & direct  &  $6$ & $1858173$ & $2.032527(23)$ & $169536$ & $2.42440(19)$ \\
         & reverse &      & $1858013$ & $2.032524(23)$ & $169625$ & $2.42457(19)$ \\ 
         & average &      &           & $2.032526(16)$ &          & $2.42449(13)$ \\ \hline
$3.6045$ & direct  &  $7$ &  $980672$ & $2.03587(4)$   &  $90172$ & $2.42370(30)$ \\
         & reverse &      &  $980802$ & $2.03579(4)$   &  $90164$ & $2.42308(29)$ \\ 
         & average &      &           & $2.03583(3)$   &          & $2.42339(21)$ \\ \hline
$3.6566$ & direct  &  $8$ &  $553830$ & $2.04160(5)$   &  $50613$ & $2.4335(5)$   \\
         & reverse &      &  $553712$ & $2.04146(5)$   &  $50613$ & $2.4343(5)$   \\
         & average &      &           & $2.04153(4)$   &          & $2.4339(4)$   \\ \hline
$3.7425$ & direct  & $10$ &  $225668$ & $2.05093(11)$  &   $3671$ & $2.4277(21)$  \\
         & reverse &      &  $225744$ & $2.05064(10)$  &   $3347$ & $2.4279(21)$  \\ 
         & average &      &           & $2.05079(7)$   &          & $2.4274(15)$  \\ \hline \hline
$3.1898$ & direct  &  $5$ & $3954338$ & $2.390473(18)$ & $358200$ & $2.98316(16)$ \\
         & reverse &      & $3954034$ & $2.390471(18)$ & $358241$ & $2.98290(16)$ \\
         & average &      &           & $2.390472(13)$ &          & $2.98303(11)$ \\ \hline
$3.2751$ & direct  &  $6$ & $1857890$ & $2.381599(30)$ & $169518$ & $2.95573(27)$ \\
         & reverse &      & $1857241$ & $2.381580(30)$ & $169508$ & $2.95597(27)$ \\
         & average &      &           & $2.381590(22)$ &          & $2.95585(19)$ \\ \hline
$3.3428$ & direct  &  $7$ &  $980305$ & $2.37987(5)$   &  $90119$ & $2.9425(4)$   \\
         & reverse &      &  $980485$ & $2.37988(5)$   &  $90135$ & $2.9431(4)$   \\
         & average &      &           & $2.37988(3)$   &          & $2.94281(29)$ \\ \hline
$3.4009$ & direct  &  $8$ &  $553957$ & $2.37832(7)$   &  $50588$ & $2.9343(6)$   \\
         & reverse &      &  $553637$ & $2.37824(7)$   &  $50588$ & $2.9356(6)$   \\
         & average &      &           & $2.37828(5)$   &          & $2.9350(5)$   \\ \hline
$3.5000$ & direct  & $10$ &  $223507$ & $2.37030(13)$  &   $2843$ & $2.901(3)$    \\
         & reverse &      &  $225641$ & $2.36994(13)$  &   $2844$ & $2.903(3)$    \\
         & average &      &           & $2.37012(9)$   &          & $2.9019(22)$    \\
\hline
\end{tabular}
\end{center}
\caption{Results for $g^2$ from the average of direct and reverse transformations with $\Delta \eta=0.0001$ and $\nqq=1000$ in $\SU(2)$ Yang-Mills theory.}
\label{tab:su2_deltaeta_0.0001}
\end{table}

\begin{table}
\begin{center}
\begin{tabular}{|c|c|c|r|l|r|l|}
\hline
$\beta$ & type & $L/a$ & $\ntraj(L)$ & \multicolumn{1}{|c|}{$g^2(L)$} & $\ntraj(2L)$ & \multicolumn{1}{|c|}{$g^2(2L)$} \\
\hline \hline
$2.9568$ & direct  &  $5$ & $3952787$ & $2.831998(25)$ & $358004$ & $3.75275(24)$ \\
         & reverse &      & $3953921$ & $2.831992(25)$ & $358027$ & $3.75288(24)$ \\
         & average &      &           & $2.831995(18)$ &          & $3.75282(17)$ \\ \hline
$3.0379$ & direct  &  $6$ & $1857290$ & $2.82828(4)$   & $169500$ & $3.7203(4)$ \\
         & reverse &      & $1857523$ & $2.82824(4)$   & $169509$ & $3.7199(4)$ \\
         & average &      &           & $2.82826(3)$   &          & $3.7201(3)$ \\ \hline
$3.0961$ & direct  &  $7$ &  $980138$ & $2.84678(7)$   &  $90110$ & $3.7332(6)$ \\
         & reverse &      &  $980150$ & $2.84693(7)$   &  $87625$ & $3.7343(7)$ \\
         & average &      &           & $2.84686(5)$   &          & $3.7337(5)$ \\ \hline
$3.1564$ & direct  &  $8$ &  $553403$ & $2.83855(10)$  &  $50545$ & $3.7025(9)$ \\
         & reverse &      &  $553398$ & $2.83851(10)$  &  $50530$ & $3.7014(10)$ \\
         & average &      &           & $2.83853(7)$   &          & $3.7019(7)$ \\ \hline
$3.2433$ & direct  & $10$ &  $225644$ & $2.85329(18)$  &   $3672$ & $3.704(4)$ \\
         & reverse &      &  $225598$ & $2.85303(18)$  &   $3060$ & $3.704(5)$ \\
         & average &      &           & $2.85316(13)$  &          & $3.704(3)$ \\ \hline \hline
$2.7124$ & direct  &  $5$ & $3622193$ & $3.56093(4)$   & $376481$ & $5.4102(5)$ \\
         & reverse &      & $3622242$ & $3.56093(4)$   & $376559$ & $5.4102(5)$ \\
         & average &      &           & $3.560933(28)$ &          & $5.4102(4)$ \\ \hline
$2.7938$ & direct  &  $6$ & $1701517$ & $3.54971(7)$   & $178344$ & $5.2909(8)$ \\
         & reverse &      & $1701666$ & $3.54968(7)$   & $178278$ & $5.2910(8)$ \\
         & average &      &           & $3.54969(5)$   &          & $5.2909(6)$ \\ \hline
$2.8598$ & direct  &  $7$ &  $898009$ & $3.54728(10)$  &  $90119$ & $5.2233(13)$ \\
         & reverse &      &  $898069$ & $3.54754(10)$  &  $90128$ & $5.2216(13)$ \\
         & average &      &           & $3.54741(7)$   &          & $5.2225(9)$ \\ \hline
$2.9115$ & direct  &  $8$ &  $507066$ & $3.56344(15)$  &  $50521$ & $5.2190(19)$ \\
         & reverse &      &  $483016$ & $3.56370(16)$  &  $50529$ & $5.2191(19)$ \\
         & average &      &           & $3.56357(11)$  &          & $5.2191(14)$ \\ \hline
$3.0071$ & direct  & $10$ &  $206723$ & $3.54904(29)$  &   $3058$ & $5.118(9)$ \\
         & reverse &      &  $196902$ & $3.54877(29)$  &   $3383$ & $5.127(9)$ \\
         & average &      &           & $3.54890(20)$  &          & $5.122(6)$ \\
\hline
\end{tabular}
\end{center}
\caption{Table~\ref{tab:su2_deltaeta_0.0001}, continued.}
\label{tab:su2_deltaeta_0.0001_bis}
\end{table}

\begin{table}
\begin{center}
\begin{tabular}{|c|r|l|l|}
\hline
$\beta$ & $L/a$ & \multicolumn{1}{|c|}{$g^2(L)$} & \multicolumn{1}{|c|}{$g^2(2L)$} \\
\hline \hline
$3.4564$ &  $5$ & $2.0371(32)$ & $2.413(15)$ \\
$3.5408$ &  $6$ & $2.0369(52)$ & $2.418(16)$ \\
$3.6045$ &  $7$ & $2.0370(55)$ & $2.397(19)$ \\
$3.6566$ &  $8$ & $2.0370(63)$ & $2.447(7)$  \\
$3.7425$ & $10$ & $2.0369(83)$ & $2.426(22)$ \\
\hline
$3.1898$ &  $5$ & $2.3800(43)$ & $2.981(23)$ \\
$3.2751$ &  $6$ & $2.3801(67)$ & $2.942(21)$ \\
$3.3428$ &  $7$ & $2.3799(67)$ & $2.968(26)$ \\
$3.4009$ &  $8$ & $2.3801(79)$ & $2.954(23)$ \\
$3.5000$ & $10$ & $2.380(11)$  & $2.870(30)$ \\
\hline
$2.9568$ &  $5$ & $2.8401(56)$ & $3.783(33)$ \\
$3.0379$ &  $6$ & $2.8401(91)$ & $3.731(35)$ \\
$3.0961$ &  $7$ & $2.840(10)$  & $3.709(31)$ \\
$3.1564$ &  $8$ & $2.840(11)$  & $3.663(34)$ \\
$3.2433$ & $10$ & $2.841(16)$  & $3.69503)$  \\
\hline
$2.7124$ &  $5$ & $3.550(10)$  & $5.456(40)$ \\
$2.7938$ &  $6$ & $3.550(14)$  & $5.287(43)$ \\
$2.8598$ &  $7$ & $3.550(15)$  & $5.310(58)$ \\
$2.9115$ &  $8$ & $3.550(16)$  & $5.168(38)$ \\
$3.0071$ & $10$ & $3.550(23)$  & $5.122(58)$ \\
\hline
\end{tabular}
\end{center}
\caption{Results for pairs of running couplings at distances $L$ and $2L$, at the same bare coupling $\beta=4/g_0^2$, reproduced from ref.~\cite[table 2]{Luscher:1992zx}.}
\label{tab:table_2_from_heplat9207010}
\end{table}

The results of this computation are reported in tables~\ref{tab:su2_deltaeta_0.0001} and~\ref{tab:su2_deltaeta_0.0001_bis}. The bare couplings and system sizes are the same that were analyzed in ref.~\cite{Luscher:1992zx} and the results obtained from our non-equilibrium Monte~Carlo calculations are fully compatible with those reported in that work, which, for the reader's convenience, we reproduce in table~\ref{tab:table_2_from_heplat9207010}. We also observe that results obtained from ``direct'' and ``reverse'' implementations of our algorithm are consistent with each other: a non-trivial check that our algorithm is correctly sampling the distribution of Euclidean action differences along the non-equilibrium trajectories. Our final results for the squared coupling at different lattice spacings are then obtained from the average of the two.

\begin{figure}[!htbp]
\begin{center}
\includegraphics*[width=\textwidth]{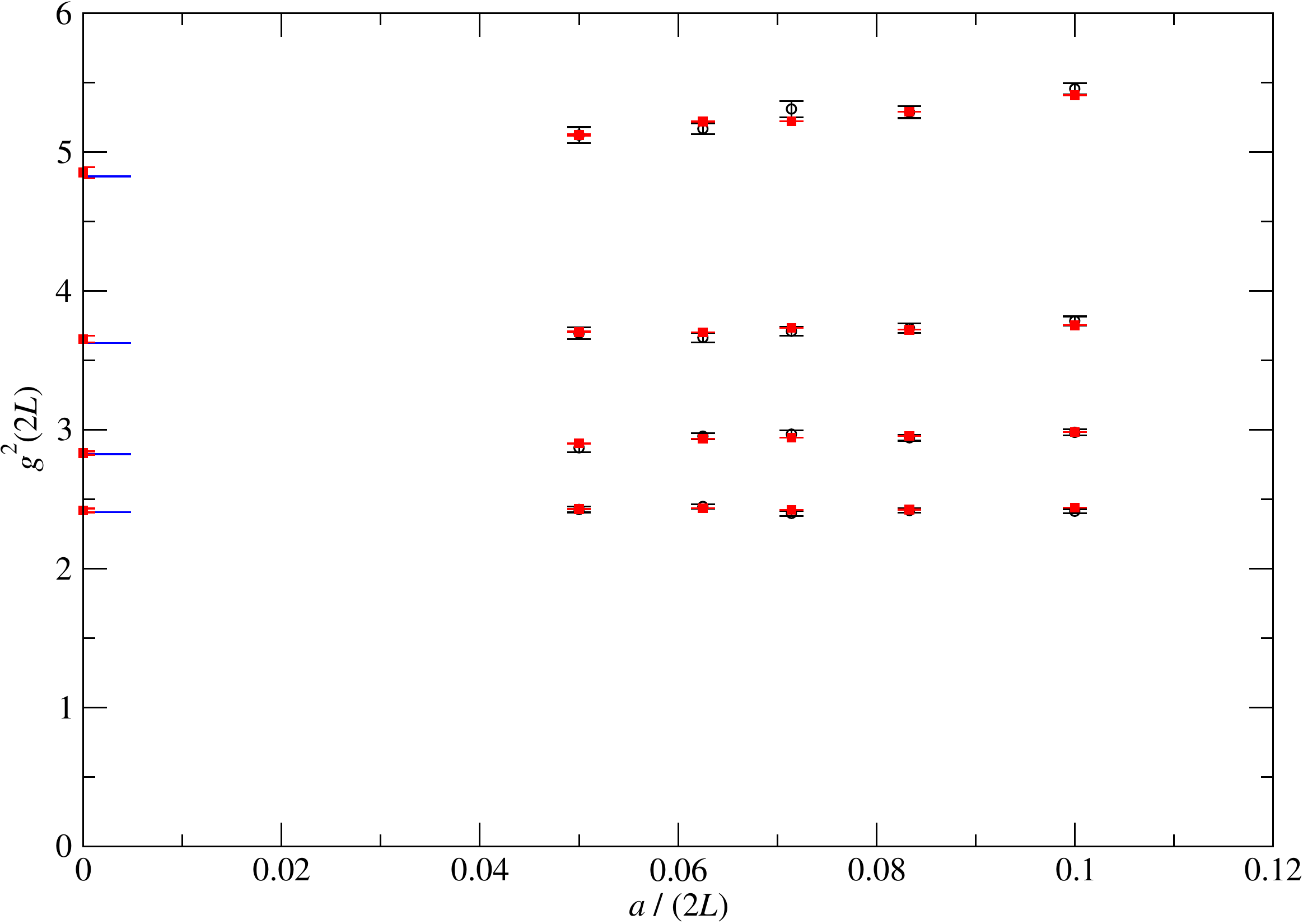}
\caption{\label{fig:SU2_coupling_at_sL_comparison} Squared $\SU(2)$ couplings evaluated at $2L$ (red squares), for four different values of $g^2(L)$, as a function of $a/(2L)$, and the corresponding continuum-extrapolated values, in comparison with the two-loop predictions (horizontal blue segments on the vertical axis). The plot also shows the results from ref.~\cite{Luscher:1992zx} (black circles).}
\end{center}
\end{figure}

In figure~\ref{fig:SU2_coupling_at_sL_comparison} we show our results for $g^2(2L)$ at different values of the lattice spacing. The data, displayed by red squares, fall on four nearly horizontal bands, corresponding to four values\footnote{These values are obtained from the averages reported in the fifth column of tables~\ref{tab:su2_deltaeta_0.0001} and~\ref{tab:su2_deltaeta_0.0001_bis} after extrapolation to the continuum limit by a constant-plus-linear-term fit in $a/L$, and are compatible with those reported in ref.~\cite{Luscher:1992zx}.} of $g^2(L)=2.059(11)$, $2.353(4)$, $2.871(14)$, and $3.546(16)$, i.e. to four values of $L$, and are plotted as a function of the lattice spacing divided by $2L$. For comparison, the figure also shows the results reported in ref.~\cite{Luscher:1992zx} as black circles. Since leading discretization effects in the lattice formulation of the Schr\"odinger functional are expected to be of order $a$, we fit each of the four data sets to the sum of a constant plus a linear function of the lattice spacing, obtaining the continuum-extrapolated results displayed by the red squares on the vertical axis of the plot: following the analysis carried out in ref.~\cite{Luscher:1992zx}, we find that all of them are very close to the two-loop perturbative predictions (horizontal blue segments on the vertical axis). One could also compare these results with three-loop perturbative predictions, which have since become available~\cite{Narayanan:1995ex, Luscher:1995nr} and which are discussed below, but at this stage we limit ourselves to note that, as was observed in ref.~\cite{Luscher:1992zx}, the two-loop perturbative predictions already provide a good approximation for the continuum-extrapolated non-perturbative results shown in figure~\ref{fig:SU2_coupling_at_sL_comparison}.

Continuum extrapolation of the results in tables~\ref{tab:su2_deltaeta_0.0001} and~\ref{tab:su2_deltaeta_0.0001_bis} reveals that the value of the squared coupling in the Schr\"odinger-functional scheme evaluated on the largest lattices is $g^2=4.85(4)$, and that the values of $g^2(2L)$ obtained in the $a \to 0$ limit from each of the four data sets are close to the value of $g^2(L)$ in the next set. As a consequence, the ratio of the corresponding length scales is very close to unity, and, following ref.~\cite{Luscher:1992zx}, can be reliably estimated using the perturbative $\beta$ function truncated at two loops. As mentioned above, in principle, the estimate of this length-scale ratio could now be refined using the three-loop perturbative $\beta$ function (or, alternatively, in a fully non-perturbative way, by additional sets of simulations), but, due to the smallness of the differences between the length scales, this would not yield significantly different results. Hence, at this step we applied exactly the same procedure that was carried out in ref.~\cite{Luscher:1992zx} (which is justified, given the exploratory nature of the present study, that does not attempt to produce results of direct phenomenological relevance), postponing the detailed discussion of the three-loop perturbative prediction and its comparison with our lattice results to the final part of this subsection.

The procedure outlined above, i.e. following the variation of the coupling through a sequence of lattices whose linear sizes are in a ratio $s$, allows one to determine the evolution of $g^2$ from the ``hadronic'' scale down to the microscopic scale, where perturbation theory becomes reliable, using the step-scaling function first introduced in ref.~\cite{Luscher:1991wu} and defined as
\begin{equation}
\label{sigma_definition}
\sigma \left( s, g^2(L) \right) = g^2(sL)
\end{equation}
and to extract the $\beta$ function of the theory from it. It is then possible to obtain the evolution of the coupling in the Schr\"odinger-functional scheme non-perturbatively over a wide range of scales, without having to perform simulations on lattices with a prohibitively large number of sites.

\begin{figure}[!htbp]
\begin{center}
\includegraphics*[width=\textwidth]{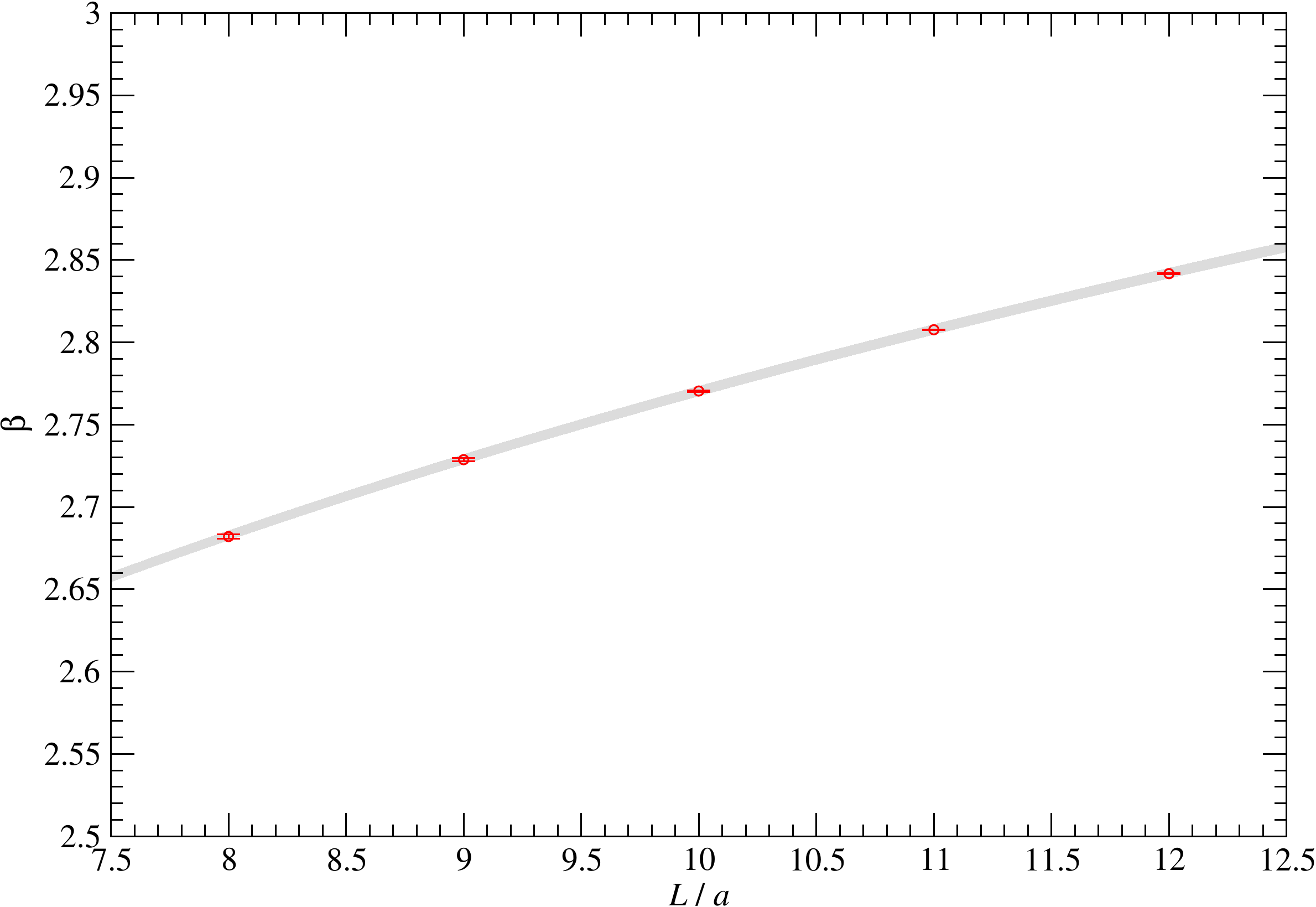}
\caption{\label{fig:SU2_coarse_lattice} The inverse-squared-bare-coupling parameter $\beta=2N/g_0^2$ corresponding to $g^2=4.85$ (red circles), as a function of $L/a$, in $\SU(2)$ Yang-Mills theory and the corresponding fitted curve $\beta=1.866(22)+0.3928(9)\cdot \ln (L/a)$, with the associated uncertainty (gray band).}
\end{center}
\end{figure}

The last step of the analysis consists, then, in the explicit construction of the $\beta$ function of the theory. To this purpose, we first focus on the low-energy regime and run an additional set of simulations for $(\beta,L/a)$ combinations yielding a value of $g^2(L)$ sufficiently close to the one extrapolated from the largest lattices listed in table~\ref{tab:su2_deltaeta_0.0001_bis} i.e. $g^2=4.85(4)$: the results are shown in table~\ref{tab:su2_coarsest}. Note that, while the number of lattice points in each direction in these lattices varies from $8$ to $12$, the corresponding lattice spacings decrease, and the Schr\"odinger-functional coupling remains nearly constant, at a value that is the largest among those that we considered for this theory. As a consequence, these lattices have nearly the same physical size, which is the largest among those that we studied for the $\SU(2)$ theory in this work. More precisely, from the values in table~\ref{tab:su2_coarsest}, we obtain the $(\beta,L/a)$ combinations corresponding to $g^2=4.85(4)$ that are listed in table~\ref{tab:su2_coarsest_couplings}. They can be fitted to the functional form $\beta=1.866(22)+0.3928(9)\cdot \ln (L/a)$, with reduced $\chi^2 \simeq 0.15$, as shown in figure~\ref{fig:SU2_coarse_lattice}.

\begin{table}
\begin{center}
\begin{tabular}{|c|c|c|r|l|}
\hline
$\beta$    & type    & $L/a$ & $\ntraj(L)$ & \multicolumn{1}{|c|}{$g^2(L)$} \\
\hline \hline
$2.68334$  & direct  &  $8$  & $136484$ & $4.8393(5)$  \\ 
           & reverse &       & $136391$ & $4.8388(5)$  \\ 
           & average &       &          & $4.8390(4)$  \\ \hline
$2.72976$  & direct  &  $9$  &  $86648$ & $4.8414(8)$  \\ 
           & reverse &       &  $86652$ & $4.8417(8)$  \\ 
           & average &       &          & $4.8416(5)$  \\ \hline
$2.77090$  & direct  & $10$  &  $55718$ & $4.8462(10)$ \\ 
           & reverse &       &  $55704$ & $4.8452(10)$ \\ 
           & average &       &          & $4.8457(7)$  \\ \hline
$2.80780$  & direct  & $11$  &  $37328$ & $4.8464(14)$ \\ 
           & reverse &       &  $37358$ & $4.8515(14)$ \\ 
           & average &       &          & $4.8489(10)$ \\ \hline
$2.84210$  & direct  & $12$  &  $26313$ & $4.8471(17)$ \\ 
           & reverse &       &  $26329$ & $4.8464(17)$ \\ 
           & average &       &          & $4.8468(12)$ \\
\hline
\end{tabular}
\end{center}
\caption{Results from the set of simulations on the lattices of the largest physical size (corresponding to $g^2 \simeq 4.85(4)$) from direct and reverse transformations with $\Delta \eta=0.0001$ and $\nqq=1000$, and their average, in $\SU(2)$ Yang-Mills theory.}
\label{tab:su2_coarsest}
\end{table}

\begin{table}
\begin{center}
\begin{tabular}{|r|l|}
\hline
\multicolumn{1}{|c|}{$L/a$} & \multicolumn{1}{|c|}{$\beta$} \\
\hline \hline
  $8$ &  $2.6820(13)$  \\
  $9$ &  $2.7287(10)$  \\
 $10$ &  $2.7704(5)$   \\
 $11$ &  $2.80767(13)$ \\
 $12$ &  $2.8417(4)$   \\
\hline
\end{tabular}
\end{center}
\caption{Couplings corresponding to $g^2=4.85$ in the $\SU(2)$ theory, as a function of $L/a$.}
\label{tab:su2_coarsest_couplings}
\end{table}

For comparison, in table~\ref{tab:table_3_from_heplat9207010} we also reproduce the analogous values obtained in ref.~\cite{Luscher:1992zx} for the lattice of largest physical size studied in that work, corresponding to $g^2=4.765$ (a value close to the squared coupling obtained from continuum extrapolation of the step-scaling function evaluated on the largest set of lattices considered therein, which reads $g^2=4.76(12)$ and is fully compatible with our result).

\begin{table}
\begin{center}
\begin{tabular}{|c|l|}
\hline
$L/a$ & \multicolumn{1}{|c|}{$\beta$} \\
\hline
\hline
~$6$ & $2.5752(28)$ \\
~$7$ & $2.6376(20)$ \\
~$8$ & $2.6957(21)$ \\
$10$ & $2.7824(22)$ \\
$12$ & $2.8485(32)$ \\
$14$ & $2.9102(62)$ \\
\hline
\end{tabular}
\end{center}
\caption{Values of $\beta=4/g_0^2$ as a function of the linear extent of the system in units of the lattice spacing, at fixed $g^2(L)=4.765$, reproduced from ref.~\cite[table 3]{Luscher:1992zx}.}
\label{tab:table_3_from_heplat9207010}
\end{table}

\begin{figure}[!htbp]
\begin{center}
\includegraphics*[width=\textwidth]{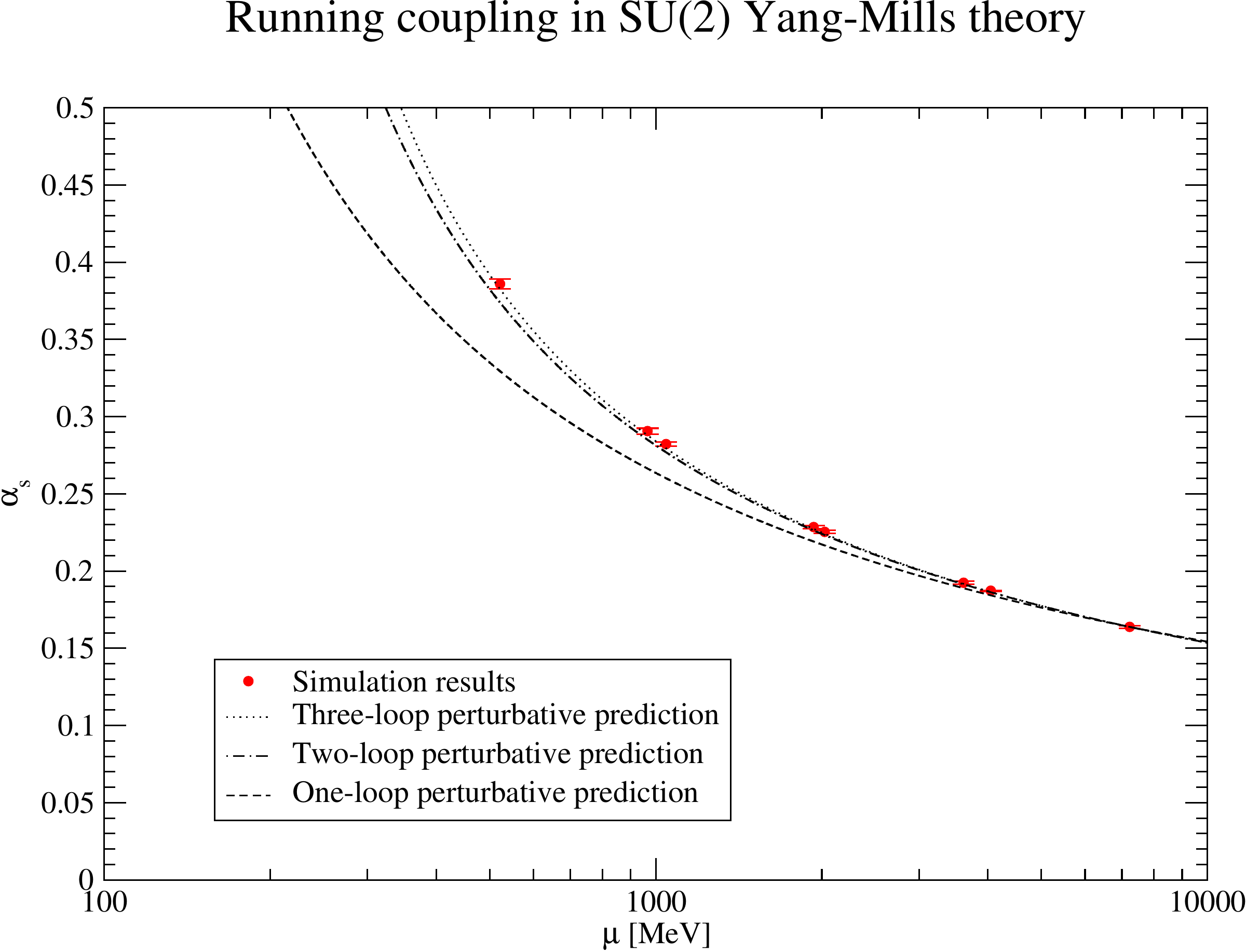}
\caption{\label{fig:su2_alphas_versus_q} The running coupling $\alpha_{\mbox{\tiny{s}}}=g^2/(4\pi)$ of $\SU(2)$ Yang-Mills theory, as a function of the momentum scale $\mu=1/L$. For comparison, the one- (dashed curve), two- (dash-dotted curve) and three-loop (dotted curve) perturbative predictions are also shown.}
\end{center}
\end{figure}

We can then make contact with a physical low-energy scale of the theory, such as the string tension $\sigma_0$, i.e. the asymptotic force between fundamental probe charges at large distance,\footnote{The phenomenological value of the string tension, extracted from Regge trajectories obtained from experimental results for mesons~\cite{Bali:2000gf}, is approximately ($440$~MeV)$^2$.} using the data reported in refs.~\cite{Luscher:1992zx, Fingberg:1992ju, Bali:1994de, Caselle:2015tza} in the range $ \beta \in [2.2,2.85]$. In particular, using the value $\sigma_0 a^2=0.00830(6)$ at $\beta=2.74$ from ref.~\cite{Bali:1994de}, one obtains that the length at which $g^2$ equals $4.85$ is $L=0.843(3)/\sqrt{\sigma_0}$, or $0.3781(14)$~fm. Thus, taking the momentum scale to be defined as $\mu=1/L$ and using the continuum extrapolations of the results listed in tables~\ref{tab:su2_deltaeta_0.0001} and~\ref{tab:su2_deltaeta_0.0001_bis}, one obtains the results for $\alphas=g^2/(4\pi)$ plotted in figure~\ref{fig:su2_alphas_versus_q}. Also shown are the analytical predictions from perturbation theory at one, two, and three loops~\cite{Gross:1973id, Politzer:1973fx, Caswell:1974gg, Jones:1974mm, Narayanan:1995ex, Luscher:1995nr}. In particular, the two-loop perturbative prediction is obtained from
\begin{equation}
\label{betafunction}
\frac{d \alphas}{d ( \ln \mu)} = - \frac{11N}{6\pi} \alphas^2  
- \frac{17N^2}{12\pi^2} \alphas^3 + O(\alphas^4),
\end{equation}
which yields
\begin{equation}
\label{integrated_beta_function}
\ln\frac{\mu_2}{\mu_1} \simeq f\left(\alphas(\mu_2)\right)-f\left(\alphas(\mu_1)\right),
\end{equation}
with
\begin{equation}
\label{f_at_two_loops}
f(x)=\frac{6 \pi}{11 N x} - \frac{51}{121}\ln \left(\frac{17 N}{22 \pi}+\frac{1}{x}\right).
\end{equation}

Note that the comparison with perturbation theory in the Schr\"odinger-functional scheme can now be pushed to three loops (which was not yet possible at the time of publication of ref.~\cite{Luscher:1992zx}): this can be done by combining the two-loop relation between the Schr\"odinger-functional coupling and the bare lattice coupling that was worked out in ref.~\cite{Narayanan:1995ex} with the one relating the bare lattice coupling to the coupling in the $\overline{\rm MS}$-scheme~\cite{Luscher:1995nr}, from which one can derive that, for the $\SU(2)$ Yang-Mills theory,
\begin{equation}
\label{two-loop_SU2_alphaMSbar_alphas}
\alphasMSbar=\alphas + c_1 \alphas^2 + c_2 \alphas^3 + O(\alphas^4).
\end{equation}
where $\alphas$ and $\alphasMSbar$ can be defined at two different momentum scales, respectively $\mu_1$ and $\mu_2$, and $c_1$ and
$c_2$ are functions of their ratio $r=\mu_2/\mu_1$:
\begin{equation}
\label{c1_and_c2_from_heplat9502001}
c_1(r)= -\frac{11}{3\pi} \ln r + 0.94327(4) ,\qquad \qquad
c_2(r)= c_1^2(r) -\frac{17}{3\pi^2} \ln r + 0.5216(5).
\end{equation}
Note that eq.~(\ref{two-loop_SU2_alphaMSbar_alphas}) can be inverted as
\begin{equation}
\label{two-loop_SU2_alphas_alphaMSbar}
\alphas = \alphasMSbar -c_1 \alpha^2_{\overline{\rm MS}} + \left( 2c_1^2 -c_2\right) \alpha^3_{\overline{\rm MS}} + O(\alpha^4_{\overline{\rm MS}}).
\end{equation}
In turn, eqs.~(\ref{two-loop_SU2_alphaMSbar_alphas}) and~(\ref{two-loop_SU2_alphas_alphaMSbar}) can be combined with the three-loop perturbative expression for the $\beta$ function in the $\overline{\rm MS}$ scheme~\cite{Tarasov:1980au} (see also refs.~\cite{Larin:1993tp, Gracey:2005vu}), which, for a purely gluonic $\SU(N)$ gauge theory, reads
\begin{equation}
\label{three-loop_MSbar_beta_function}
\frac{d \alphasMSbar}{d ( \ln \mu)} = -\frac{11N}{6\pi}\alpha^2_{\overline{\rm MS}} -\frac{17N^2}{12\pi^2}\alpha^3_{\overline{\rm MS}} -\frac{2857N^3}{1728\pi^3}\alpha^4_{\overline{\rm MS}} +O ( \alpha^5_{\overline{\rm MS}} ),
\end{equation}
to obtain the three-loop perturbative $\beta$ function for the $\SU(2)$ Yang-Mills theory in the Schr\"odinger-functional scheme:
\begin{equation}
\label{three-loop_SU2_SF_beta_function}
\frac{d \alphas}{d ( \ln \mu)} = -\frac{11}{3\pi}\alphas^2 -\frac{17}{3\pi^2}\alphas^3 -\left[ \frac{2857}{216\pi^3} +\frac{17}{3\pi^2}c_1 +\frac{11}{3\pi}(c_1^2-c_2) \right]\alphas^4 +O ( \alphas^5 ).
\end{equation}
Integrating eq.~(\ref{three-loop_SU2_SF_beta_function}) (assuming $r=1$), one obtains again an expression like the one given in eq.~(\ref{integrated_beta_function}), but with eq.~(\ref{f_at_two_loops}) replaced by
\begin{equation}
\label{f_at_three_loops}
f(x)=-\frac{b_1}{2b_0^2} \ln \left( \frac{x^2}{b_0+b_1x+b_2x^2} \right) - \frac{1}{b_0x} + \frac{b_1^2-2b_0 b_2}{b_0^2\sqrt{-\Delta}} \arctan \left( \frac{b_1+2b_2x}{\sqrt{-\Delta}} \right),
\end{equation}
having defined
\begin{equation}
\label{three-loop_SU2_beta_function_coefficients}
b_0=-\frac{11}{3\pi},\quad b_1=-\frac{17}{3\pi^2} ,\quad b_2=-\left[ \frac{2857}{216\pi^3} +\frac{17}{3\pi^2}c_1 +\frac{11}{3\pi}(c_1^2-c_2) \right],\quad \mbox{and } \Delta=b_1^2-4 b_0 b_2.
\end{equation}

Our numerical results are in very good agreement with those reported in ref.~\cite{Luscher:1992zx}, and confirm the accuracy of the three-loop (and two-loop) perturbative $\beta$ function down to $\mu \sim 1$~GeV. For a comparison, in table~\ref{tab:table_5_from_heplat9207010} we reproduce the values of the running coupling in the Schr\"odinger-functional scheme, as a function of $L/L_8$ (with $L_8$ denoting the length scale at which $g^2(L)$ reaches the largest value studied in that work), that were obtained in ref.~\cite{Luscher:1992zx}. Note that the largest value of $g^2$ that was considered in ref.~\cite{Luscher:1992zx} (namely $4.765$) is close but not exactly equal to ours (which is $4.85(4)$). As remarked above, this mismatch arises at an intermediate step in the calculation, in particular at the level of the continuum extrapolation of the results for $g^2(2L)$, which (as shown for instance in fig.~\ref{fig:SU2_coupling_at_sL_comparison}) are affected by different statistical uncertainties in the two studies. Note, however, that the $g^2$ value that was obtained from the continuum extrapolation of the step-scaling function on the set of largest lattices in ref.~\cite{Luscher:1992zx} is $4.76(12)$, in perfect agreement with our value within one standard deviation. The overall agreement between the two works remains very good.

\begin{table}
\begin{center}
\begin{tabular}{|l|l|}
\hline
\multicolumn{1}{|c|}{$L/L_8$} & \multicolumn{1}{|c|}{$g^2(L)$} \\
\hline
\hline
$1$         &  $4.765$ \\
$0.500(23)$ &  $3.550$ \\
$0.249(19)$ &  $2.840$ \\
$0.124(13)$ &  $2.380$ \\
$0.070(8)$  &  $2.037$ \\
\hline
\end{tabular}
\end{center}
\caption{Values of the running coupling as a function of $L$, in units of the length scale $L_8$, such that $g^2(L_8)=4.765$, reproduced from ref.~\cite[table 5]{Luscher:1992zx}.}
\label{tab:table_5_from_heplat9207010}
\end{table}

At the three lowest energy scales that we studied, the two-loop perturbative prediction systematically underestimates the non-perturbative Monte~Carlo results, while the three-loop perturbative prediction is in agreement with them. As a curiosity, extrapolating our results to high energies using the two-loop (or the three-loop) perturbative $\beta$ function, at the pole mass of the physical $Z^0$ boson of the Standard Model we obtain $\alphas\left(m_{Z^0}\right) = 0.1081(6)$ in the Schr\"odinger-functional scheme for the purely gluonic $\SU(2)$ Yang-Mills theory.

\subsection{Results for the $\SU(3)$ theory}
\label{subsec:results_for_the_SU3_theory}

Our study of the $\SU(3)$ theory follows closely the one presented in subsection~\ref{subsec:results_for_the_SU2_theory} for the $N=2$ case. In this case, we compare our results with those reported in ref.~\cite{Luscher:1993gh}. The main difference with respect to the $\SU(2)$ case is that the rank of the algebra of group generators is $2$ (instead of $1$) and the fundamental domain is specified by the two parameters $\eta$ and $\nu$ (instead of just $\eta$). We run our non-equilibrium simulations around $\eta=0$ (instead of $\eta=\pi/4$), at fixed $\nu=0$. In addition, as mentioned earlier, for the $\SU(3)$ theory we set $c_t$, the improvement coefficient for plaquettes parallel to the Euclidean-time direction touching the boundaries, to $1-0.089 g_0^2$ (instead of $1$). As before, 
we study the running coupling in the Schr\"odinger-functional scheme through simulations on lattices of physical linear sizes $L$ and $sL$, where $s=2$ throughout, except for the last set of four simulations, corresponding to $g^2(L)=2.77$, for which $s=3/2$. We emphasize that this choice is simply due to our decision of reproducing the exact choice of parameters as in ref.~\cite{Luscher:1993gh}, for an easier comparison with that reference work, and is not due to any inherent limitation of our algorithm.

\begin{table}
\begin{center}
\begin{tabular}{|c|c|c|r|l|c|r|l|}
\hline
$\beta$ & type & $L/a$ & $\ntraj(L)$ & \multicolumn{1}{|c|}{$g^2(L)$} & $sL/a$ & $\ntraj(sL)$ & \multicolumn{1}{|c|}{$g^2(sL)$} \\
\hline \hline
$8.7522$ & direct  &  $5$ & $186628$ & $1.24538(4)$   & $10$ & $65614$ & $1.42973(9)$  \\
         & reverse &      & $186709$ & $1.24532(4)$   &      & $65612$ & $1.42974(9)$  \\
         & average &      &          & $1.245352(26)$ &      &         & $1.42974(7)$  \\
\hline
$8.8997$ & direct  &  $6$ &  $86973$ & $1.24744(11)$  & $12$ & $31061$ & $1.43280(16)$ \\
         & reverse &      &  $86955$ & $1.24809(12)$  &      & $31064$ & $1.43316(16)$ \\
         & average &      &          & $1.24777(8)$   &      &         & $1.43298(11)$ \\
\hline
$9.035$  & direct  &  $7$ &  $45757$ & $1.24720(21)$  & $14$ & $16127$ & $1.43212(26)$ \\
         & reverse &      &  $45737$ & $1.2516(3)$    &      & $16127$ & $1.43190(26)$ \\
         & average &      &          & $1.24938(20)$  &      &         & $1.43201(18)$ \\
\hline
$9.1544$ & direct  &  $8$ &  $26186$ & $1.2475(4)$    & $16$ &  $9214$ & $1.4296(4)$   \\
         & reverse &      &  $26195$ & $1.2464(3)$    &      &  $9214$ & $1.4298(4)$   \\
         & average &      &          & $1.24692(26)$  &      &         & $1.42972(27)$ \\
\hline \hline
$8.1555$ & direct  &  $5$ & $186736$ & $1.43513(3)$   & $10$ & $65609$ & $1.69371(13)$ \\
         & reverse &      & $186641$ & $1.43514(3)$   &      & $65622$ & $1.69371(13)$ \\
         & average &      &          & $1.435138(23)$ &      &         & $1.69371(9)$  \\
\hline
$8.3124$ & direct  &  $6$ &  $86990$ & $1.43280(5)$   & $12$ & $31051$ & $1.69354(22)$ \\
         & reverse &      &  $87013$ & $1.43284(5)$   &      & $31063$ & $1.69341(22)$ \\
         & average &      &          & $1.43282(4)$   &      &         & $1.69348(15)$ \\
\hline
$8.4442$ & direct  &  $7$ &  $45748$ & $1.43285(11)$  & $14$ & $16126$ & $1.6934(4)$   \\
         & reverse &      &  $45756$ & $1.43291(12)$  &      & $16127$ & $1.6933(3)$   \\
         & average &      &          & $1.43288(8)$   &      &         & $1.69334(25)$ \\
\hline
$8.5598$ & direct  &  $8$ &  $26178$ & $1.43418(28)$  & $16$ &  $9215$ & $1.6925(5)$   \\
         & reverse &      &  $26177$ & $1.43705(36)$  &      &  $9214$ & $1.6932(5)$   \\
         & average &      &          & $1.43561(23)$  &      &         & $1.6929(4)$   \\
\hline
\end{tabular}
\end{center}
\caption{Results for $g^2(L)$ and $g^2(sL)$ from direct and reverse transformations with $\Delta \eta=0.0001$ and $\nqq=1000$ in $\SU(3)$ Yang-Mills theory, and their average.}
\label{tab:su3_deltaeta_0.0001}
\end{table}

\begin{table}
\begin{center}
\begin{tabular}{|c|c|c|r|l|c|r|l|}
\hline
$\beta$ & type & $L/a$ & $\ntraj(L)$ & \multicolumn{1}{|c|}{$g^2(L)$} & $sL/a$ & $\ntraj(sL)$ & \multicolumn{1}{|c|}{$g^2(sL)$} \\
\hline \hline
$7.5687$ & direct  &  $5$ & $186698$ & $1.69576(4)$  & $10$ & $65614$ & $2.08488(19)$ \\
         & reverse &      & $186669$ & $1.69575(4)$  &      & $65610$ & $2.08457(19)$ \\
         & average &      &          & $1.69576(3)$  &      &         & $2.08473(13)$ \\
\hline
$7.717$  & direct  &  $6$ &  $86884$ & $1.69729(7)$  & $12$ & $31059$ & $2.0899(3)$   \\
         & reverse &      &  $86954$ & $1.69727(7)$  &      & $31055$ & $2.0897(3)$   \\
         & average &      &          & $1.69728(5)$  &      &         & $2.08978(23)$ \\
\hline
$7.8521$ & direct  &  $7$ &  $45754$ & $1.69457(11)$ & $14$ & $16128$ & $2.0864(5)$   \\
         & reverse &      &  $45750$ & $1.69471(11)$ &      & $16128$ & $2.0861(5)$   \\
         & average &      &          & $1.69464(8)$  &      &         & $2.0862(4)$   \\
\hline
$7.9741$ & direct  &  $8$ &  $26170$ & $1.69156(22)$ & $16$ &  $9215$ & $2.0797(8)$   \\
         & reverse &      &  $26179$ & $1.69174(22)$ &      &  $9215$ & $2.0794(8)$   \\
         & average &      &          & $1.69165(16)$ &      &         & $2.0795(5)$   \\
\hline
$8.165$  & direct  & $10$ &  $10358$ & $1.6934(4)$   & $20$ &  $3456$ & $2.0776(15)$  \\
         & reverse &      &  $10360$ & $1.6944(5)$   &      &  $3456$ & $2.0801(15)$  \\
         & average &      &          & $1.6939(3)$   &      &         & $2.0788(11)$  \\
\hline \hline
$6.9671$ & direct  &  $5$ & $186610$ & $2.10173(6)$  & $10$ & $65604$ & $2.7815(3)$   \\
         & reverse &      & $186666$ & $2.10174(6)$  &      & $65597$ & $2.7819(3)$   \\
         & average &      &          & $2.10174(4)$  &      &         & $2.78173(23)$ \\
\hline
$7.1214$ & direct  &  $6$ &  $86898$ & $2.09911(11)$ & $12$ & $31045$ & $2.7779(5)$   \\
         & reverse &      &  $86962$ & $2.09915(11)$ &      & $31054$ & $2.7784(6)$   \\
         & average &      &          & $2.09913(8)$  &      &         & $2.7781(4)$   \\
\hline
$7.2549$ & direct  &  $7$ &  $45756$ & $2.09585(17)$ & $14$ & $16128$ & $2.7723(9)$   \\
         & reverse &      &  $45735$ & $2.09610(17)$ &      & $16128$ & $2.7694(9)$   \\
         & average &      &          & $2.09597(12)$ &      &         & $2.7708(6)$   \\
\hline
$7.3632$ & direct  &  $8$ &  $26141$ & $2.10007(25)$ & $16$ &  $9215$ & $2.7754(13)$  \\
         & reverse &      &  $26176$ & $2.10032(25)$ &      &  $9215$ & $2.7764(13)$  \\
         & average &      &          & $2.10020(17)$ &      &         & $2.7759(9)$   \\
\hline
$7.5525$ & direct  & $10$ &  $10355$ & $2.1052(6)$   & $20$ &  $3456$ & $2.7654(26)$  \\
         & reverse &      &  $10357$ & $2.1016(5)$   &      &  $3456$ & $2.7726(28)$  \\
         & average &      &          & $2.1034(4)$   &      &         & $2.7690(19)$  \\
\hline
\end{tabular}
\end{center}
\caption{Table~\ref{tab:su3_deltaeta_0.0001}, continued.}
\label{tab:su3_deltaeta_0.0001_bis}
\end{table}

\begin{table}
\begin{center}
\begin{tabular}{|c|c|c|r|l|c|r|l|}
\hline
$\beta$ & type & $L/a$ & $\ntraj(L)$ & \multicolumn{1}{|c|}{$g^2(L)$} & $sL/a$ & $\ntraj(sL)$ & \multicolumn{1}{|c|}{$g^2(sL)$} \\
\hline \hline
$6.5512$ & direct  &  $6$ & $86884$ & $2.76996(18)$ & $~9$ & $101295$ & $3.4801(4)$   \\
         & reverse &      & $86862$ & $2.76947(18)$ &      & $101282$ & $3.4803(4)$   \\
         & average &      &         & $2.76971(13)$ &      &          & $3.48020(27)$ \\
\hline
$6.786$  & direct  &  $8$ & $26176$ & $2.7766(4)$   & $12$ &  $31053$ & $3.4812(9)$   \\
         & reverse &      & $26166$ & $2.7756(4)$   &      &  $31049$ & $3.4808(9)$   \\
         & average &      &         & $2.77609(30)$ &      &          & $3.4810(6)$   \\
\hline
$6.9748$ & direct  & $10$ & $10362$ & $2.7728(8)$   & $15$ &  $12096$ & $3.4646(18)$  \\
         & reverse &      & $10359$ & $2.7719(8)$   &      &  $12096$ & $3.4637(17)$  \\
         & average &      &         & $2.7723(6)$   &      &          & $3.4641(12)$  \\
\hline
$7.119$  & direct  & $12$ &  $4894$ & $2.8201(25)$  & $18$ &   $5751$ & $3.477(3)$    \\
         & reverse &      &  $4894$ & $2.8238(27)$  &      &   $5753$ & $3.478(3)$    \\
         & average &      &         & $2.8219(18)$  &      &          & $3.4774(22)$  \\
\hline
\end{tabular}
\end{center}
\caption{Table~\ref{tab:su3_deltaeta_0.0001_bis}, continued.}
\label{tab:su3_deltaeta_0.0001_ter}
\end{table}

\begin{figure}[!htbp]
\begin{center}
\includegraphics*[width=\textwidth]{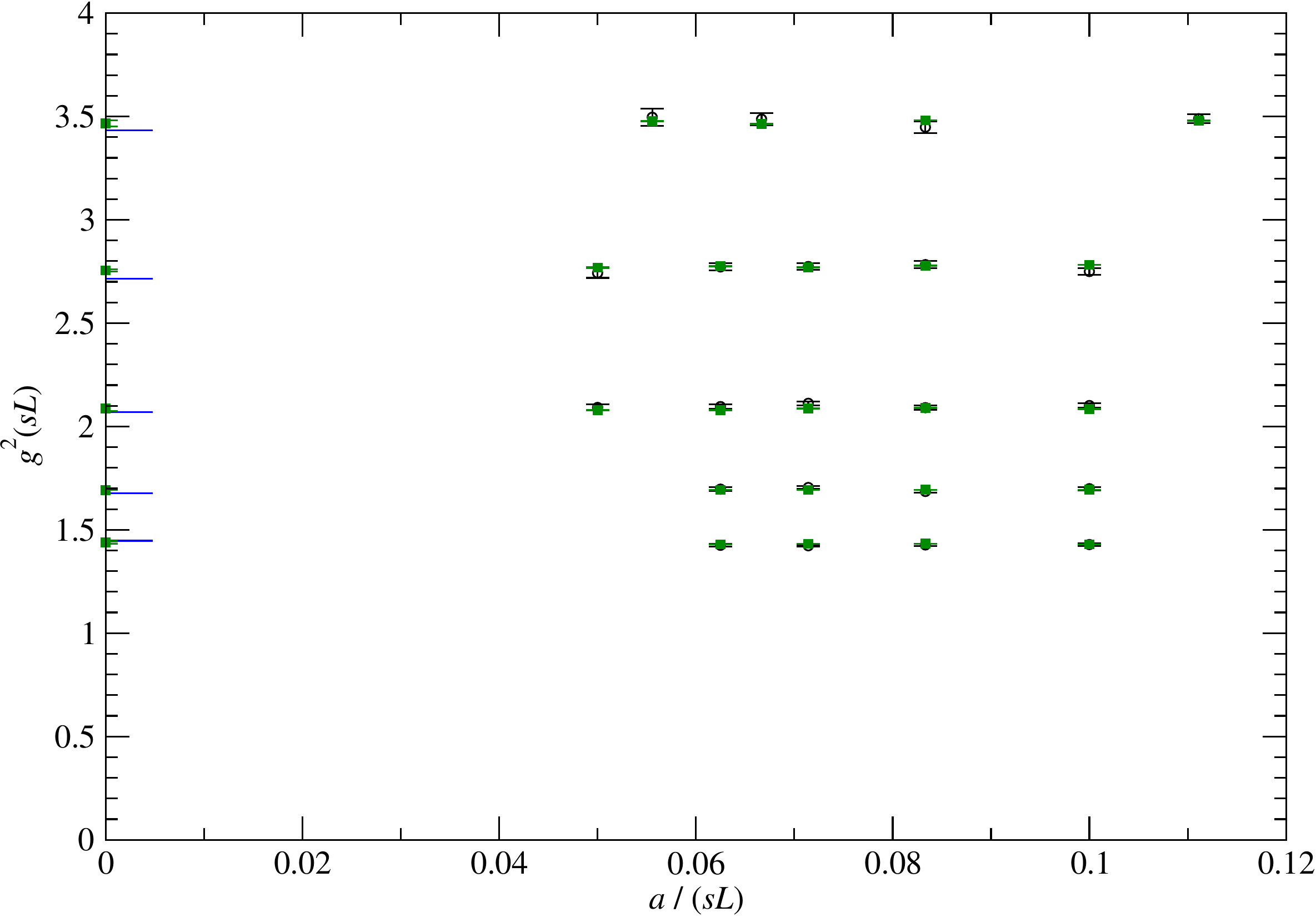}
\caption{\label{fig:SU3_coupling_at_sL_comparison} Squared $\SU(3)$ couplings evaluated at $sL$ (green squares) for five different values of $g^2(L)$ (from bottom to top: $g^2(L)=1.2571(32)$, $1.4252(34)$, $1.6943(47)$, $2.0902(51)$, and $2.793(17)$) and for $s=2$ in all cases except for $g^2(L)=2.793(17)$, for which $s=3/2$. The data are plotted against $a/(sL)$. On the vertical axis, the continuum-extrapolated values (green symbols) are compared with the two-loop predictions (horizontal blue segments). The figure also displays the results from ref.~\cite{Luscher:1993gh} as black circles.}
\end{center}
\end{figure}

Our results, reported in tables~\ref{tab:su3_deltaeta_0.0001},~\ref{tab:su3_deltaeta_0.0001_bis}, and~\ref{tab:su3_deltaeta_0.0001_ter}, are plotted against $a/(sL)$ in figure~\ref{fig:SU3_coupling_at_sL_comparison}, which also shows their extrapolation to the continuum limit, and the comparison with the two-loop perturbative predictions.

For comparison, we also reproduce the results obtained with a conventional Monte~Carlo algorithm in ref.~\cite{Luscher:1993gh} in table~\ref{tab:table_2_from_heplat9309005}.

\begin{table}
\begin{center}
\begin{tabular}{|l|r|l|l|}
\hline
\multicolumn{1}{|c|}{$\beta$} & \multicolumn{1}{|c|}{$L/a$} & \multicolumn{1}{|c|}{$g^2(L)$} & \multicolumn{1}{|c|}{$g^2(sL)$} \\
\hline
\hline
$8.7522$ &  $5$ & $1.2430(12)$ & $1.4284(53)$ \\
$8.8997$ &  $6$ & $1.2430(13)$ & $1.4270(50)$ \\
$9.035$  &  $7$ & $1.2430(15)$ & $1.4230(50)$ \\
$9.1544$ &  $8$ & $1.2430(14)$ & $1.4250(58)$ \\
\hline
$8.1555$ &  $5$ & $1.4300(21)$ & $1.6986(71)$ \\
$8.3124$ &  $6$ & $1.4300(20)$ & $1.6859(73)$ \\
$8.4442$ &  $7$ & $1.4300(18)$ & $1.7047(77)$ \\
$8.5598$ &  $8$ & $1.4300(21)$ & $1.6966(90)$ \\
\hline
$7.5687$ &  $5$ & $1.6950(26)$ & $2.101(11)$  \\
$7.717$  &  $6$ & $1.6950(26)$ & $2.091(11)$  \\
$7.8521$ &  $7$ & $1.6950(28)$ & $2.112(10)$  \\
$7.9741$ &  $8$ & $1.6950(28)$ & $2.096(11)$  \\
$8.165$  & $10$ & $1.6950(31)$ & $2.092(15)$  \\
\hline
$6.9671$ &  $5$ & $2.1000(41)$ & $2.750(17)$  \\
$7.1214$ &  $6$ & $2.1000(39)$ & $2.783(18)$  \\
$7.2549$ &  $7$ & $2.1000(43)$ & $2.774(16)$  \\
$7.3632$ &  $8$ & $2.1000(45)$ & $2.772(17)$  \\
$7.5525$ & $10$ & $2.1000(42)$ & $2.743(24)$  \\
\hline
$6.5512$ &  $6$ & $2.7700(69)$ & $3.489(22)$  \\
$6.786$  &  $8$ & $2.7700(73)$ & $3.448(28)$  \\
$6.9748$ & $10$ & $2.7700(76)$ & $3.487(29)$  \\
$7.119$  & $12$ & $2.770(11)$  & $3.496(42)$  \\
\hline
\end{tabular}
\end{center}
\caption{Evolution of the renormalized coupling at different length scales, reproduced from ref.~\cite[table 2]{Luscher:1993gh}. $s=2$ for the first four sets of data, while $s=3/2$ for the last data set, corresponding to $g^2(L) \simeq 2.77$.}
\label{tab:table_2_from_heplat9309005}
\end{table}

\begin{figure}[!htbp]
\begin{center}
\includegraphics*[width=\textwidth]{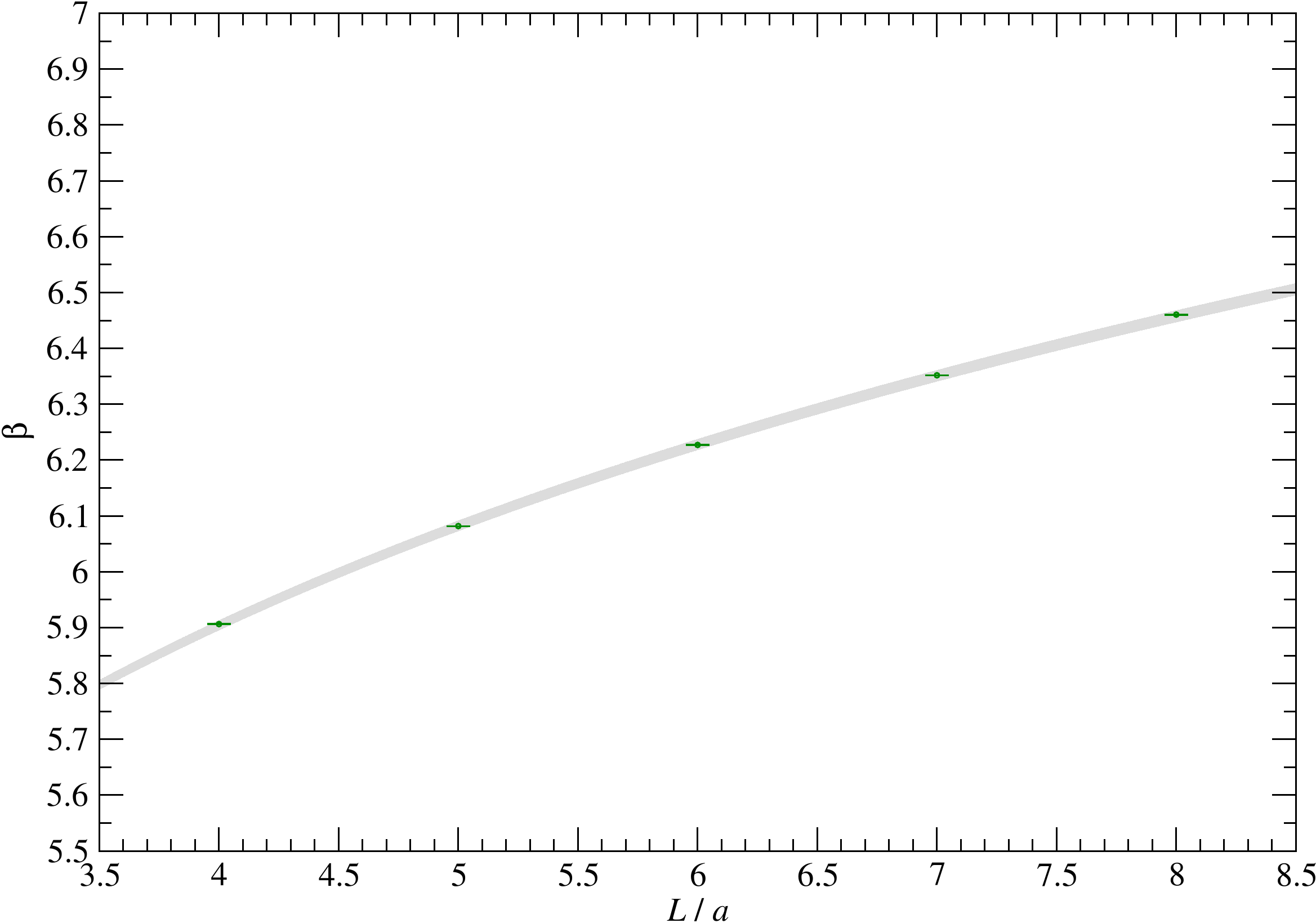}
\caption{\label{fig:SU3_coarse_lattice} The inverse-squared-bare-coupling parameter $\beta=2N/g_0^2$ corresponding to $g^2=3.467$ (green circles), as a function of $L/a$, in $\SU(3)$ Yang-Mills theory and the corresponding fitted curve $\beta=4.797(6)+0.798(4)\cdot\ln(L/a)$, with the associated uncertainty (gray band).}
\end{center}
\end{figure}

As for the analysis of the $N=2$ theory, we focus on the value of the squared coupling on the largest system that we simulated, which in this case is $g^2=3.467(15)$. We therefore run an additional set of simulations that approximately correspond to this value of the squared coupling, for the $(\beta,L/a)$ combinations listed in table~\ref{tab:su3_coarsest}. Then, we estimate the actual values of $\beta$ that would yield $g^2=3.467$ using the two-loop approximation in eq.~(\ref{integrated_beta_function}) and the parametrization of the lattice spacing as a function of $\beta$ reported in ref.~\cite[eq.~(2.6)]{Necco:2001xg}, which holds in the whole range of $\beta$ values reported in table~\ref{tab:su3_coarsest}, and take the difference with respect to the simulated values of $\beta$ as an estimate of the uncertainty on the results. This leads to the values listed in table~\ref{tab:su3_coarsest_couplings}, which can be fitted to the functional form $\beta=4.797(6)+0.798(4)\cdot\ln(L/a)$; both the data points and the fitted curve are plotted in figure~\ref{fig:SU3_coarse_lattice}.

\begin{table}
\begin{center}
\begin{tabular}{|c|c|c|r|l|}
\hline
$\beta$    & type    & $L/a$ & $\ntraj(L)$ & \multicolumn{1}{|c|}{$g^2(L)$} \\
\hline \hline
$5.90603$ & direct  & $4$  & $136959$ & $3.46816(17)$ \\
          & reverse &      & $136974$ & $3.46808(17)$ \\
          & average &      &          & $3.46812(12)$ \\ \hline
$6.0818$  & direct  & $5$  &  $53615$ & $3.4661(3)$   \\
          & reverse &      &  $53534$ & $3.4671(3)$   \\
          & average &      &          & $3.46664(22)$ \\ \hline
$6.2274$  & direct  & $6$  &  $24926$ & $3.4661(5)$   \\
          & reverse &      &  $24931$ & $3.4658(5)$   \\
          & average &      &          & $3.4659(4)$   \\ \hline
$6.35198$ & direct  & $7$  &  $13140$ & $3.4658(8)$   \\
          & reverse &      &  $13146$ & $3.4666(8)$   \\
          & average &      &          & $3.4662(6)$   \\ \hline
$6.46093$ & direct  & $8$  &   $7474$ & $3.4662(13)$  \\
          & reverse &      &   $7475$ & $3.4647(12)$  \\
          & average &      &          & $3.4655(9)$   \\
\hline
\end{tabular}
\end{center}
\caption{Results of our simulations on the largest lattice (corresponding to $g^2 \simeq 3.467(15)$) from direct and reverse transformations with $\Delta \eta=0.0001$ and $\nqq=1000$, and their average, in $\SU(3)$ Yang-Mills theory.}
\label{tab:su3_coarsest}
\end{table}

\begin{table}
\begin{center}
\begin{tabular}{|c|l|}
\hline
$L/a$ & \multicolumn{1}{|c|}{$\beta$} \\
\hline \hline
  $4$ &  $5.90632(28)$  \\
  $5$ &  $6.08169(11)$  \\
  $6$ &  $6.22702(38)$  \\
  $7$ &  $6.35168(30)$  \\
  $8$ &  $6.46035(58)$  \\
\hline
\end{tabular}
\end{center}
\caption{Couplings corresponding to $g^2=3.467$ in the $\SU(3)$ theory, as a function of $L/a$.}
\label{tab:su3_coarsest_couplings}
\end{table}

For comparison, in table~\ref{tab:table_3_from_heplat9309005} we also reproduce the analogous results from ref.~\cite[table 3]{Luscher:1993gh}, which correspond to a similar value of the squared coupling: $g^2(L) \simeq 3.48$.

\begin{table}
\begin{center}
\begin{tabular}{|c|l|}
\hline
$L/a$ & \multicolumn{1}{|c|}{$\beta$} \\
\hline \hline
$~4$ &  $5.9044(38)$ \\
$~5$ &  $6.0829(35)$ \\
$~6$ &  $6.2204(25)$ \\
$~7$ &  $6.3443(43)$ \\
$~8$ &  $6.4527(46)$ \\
$~9$ &  $6.5539(80)$ \\
$12$ &  $6.775(13)$  \\
$15$ &  $6.973(10)$  \\
\hline
\end{tabular}
\end{center}
\caption{Couplings corresponding to $g^2(L) \simeq 3.48$ in the $\SU(3)$ theory, as a function of $L/a$, reproduced from ref.~\cite[table 3]{Luscher:1993gh}.}
\label{tab:table_3_from_heplat9309005}
\end{table}

\begin{figure}[!htbp]
\begin{center}
\includegraphics*[width=\textwidth]{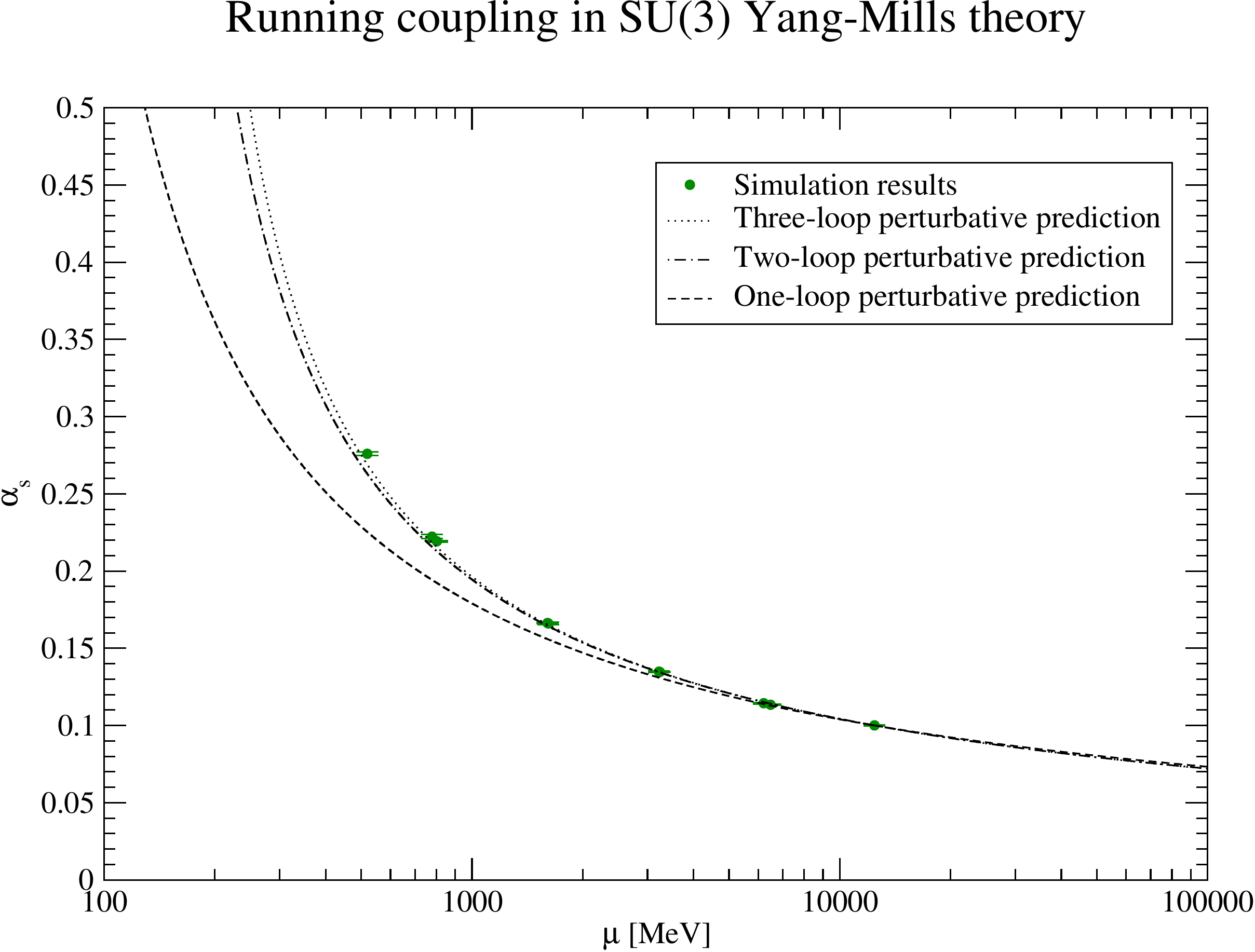}
\caption{\label{fig:su3_alphas_versus_q} Running coupling $\alpha_{\mbox{\tiny{s}}}=g^2/(4\pi)$ in $\SU(3)$ Yang-Mills theory, plotted against the momentum scale $\mu=1/L$. The dashed, dash-dotted, and dotted curves represent the perturbative predictions at one, at two, and at three loops, respectively.}
\end{center}
\end{figure}

Finally, we can then match our low-energy results with a physical scale of the theory. In this case, we choose Sommer's parameter $r_0$, defined as the distance $r$ at which the force $F$ between fundamental probe charges satisfies $r^2 F(r)=1.65$~\cite{Sommer:1993ce}. Using the high-precision parametrization of the lattice spacing in units of $r_0$ that was reported in ref.~\cite{Necco:2001xg}:
\begin{equation}
\ln\frac{a}{r_0} =-1.6804-1.7331\cdot(\beta - 6) +0.7849\cdot(\beta - 6)^2 -0.4428\cdot(\beta - 6)^3, \quad \mbox{for } 5.7 \le \beta \le 6.92,
\end{equation}
we deduce that the value of the lattice spacing at $\beta=6.08169$ equals $a=0.1625157\cdot r_0$, and that, as a consequence, $L=5a=0.8125786\cdot r_0$. We can convert this into physical units by taking the estimate for the physical value of $r_0$ in QCD $r_0=0.468(4)$~fm from ref.~\cite{Bazavov:2011nk}\footnote{This value is consistent with other estimates, such as $r_0 = 0.469(7)$~fm from ref.~\cite{Gray:2005ur}, or $r_0 = 0.462(11)(4)$~fm from ref.~\cite{Aubin:2004wf}.}, which leads to $L=0.3803(33)$~fm and $q=1/L=0.519(4)$~GeV. Proceeding as for the $\SU(2)$ theory, we can then compare our lattice results with perturbative predictions. For the theory with $N=3$ color charges, the perturbative three-loop $\beta$ function in the Schr\"odinger-functional scheme was worked out in ref.~\cite{Bode:1998hd} and reads
\begin{equation}
\label{three-loop_SU3_SF_beta_function}
\frac{d \alphas}{d ( \ln \mu)} = -\frac{11}{2\pi}\alphas^2 -\frac{51}{4\pi^2}\alphas^3 -0.966(18)\alphas^4 +O ( \alphas^5 ).
\end{equation}
Integrating eq.~(\ref{three-loop_SU3_SF_beta_function}), we finally obtain the curves plotted in figure~\ref{fig:su3_alphas_versus_q} alongside our simulation results.

Also in this case, we can compare our results with those from conventional Monte~Carlo simulations in ref.~\cite{Luscher:1993gh}: table~\ref{tab:table_5_from_heplat9309005} reproduces the values for $g^2(L)$ that were obtained in that work at different length scales $L$, in units of $L_{\mbox{\tiny{max}}}$, defined as the box size at which $g^2 \simeq 3.48$.

\begin{table}
\begin{center}
\begin{tabular}{|l|l|}
\hline
\multicolumn{1}{|c|}{$L/L_{\mbox{\tiny{max}}}$} & \multicolumn{1}{|c|}{$g^2(L)$} \\
\hline
\hline
$1$         &  $3.480$ \\
$0.664(19)$ &  $2.770$ \\
$0.332(14)$ &  $2.100$ \\
$0.165(9)$  &  $1.695$ \\
$0.084(6)$  &  $1.430$ \\
$0.040(4)$  &  $1.243$ \\
\hline
\end{tabular}
\end{center}
\caption{Values of the running coupling as a function of $L$, in units of the length scale $L_{\mbox{\tiny{max}}}$, defined by the condition $g^2(L_{\mbox{\tiny{max}}})=3.48$, reproduced from ref.~\cite[table 5]{Luscher:1993gh}.}
\label{tab:table_5_from_heplat9309005}
\end{table}

While the values of the squared Schr\"odinger-functional coupling corresponding to the lattice with the largest physical size obtained in our work and in ref.~\cite{Luscher:1993gh} are slightly different (a mismatch due to the different level of precision of the extrapolations involved), we see that the behavior of the running coupling obtained in the two works with two radically different numerical algorithms is quantitatively consistent. We conclude that, as in the $N=2$ case, our numerical results reproduce those obtained from ``conventional'' (equilibrium) Monte~Carlo calculations~\cite{Luscher:1993gh}. Our results are in very good quantitative agreement with the analytical predictions from the three-loop perturbative $\beta$ function, eq.~(\ref{three-loop_SU3_SF_beta_function}) (and also with its truncation at two loops): this holds for all values of $\mu$, down to $\mu \sim 1$~GeV. In this theory, an extrapolation of $\alphas$ to the pole mass of the $Z^0$ boson of the Standard Model yields $\alphas\left(m_{Z^0}\right) = 0.07297(19)$. When converted to the modified minimal-subtraction ($\overline{\rm MS}$) scheme via the one-loop relation $\alphasMSbar=\alphas+1.25563(4)\alphas^2$~\cite{Luscher:1993gh}, this corresponds to approximately $0.07966(22)$. For comparison, note that at this scale the value of $\alphas$ in the modified minimal-subtraction scheme recently reported in ref.~\cite{Bruno:2017gxd} for QCD with five quark flavors, combining the three-flavor running with a perturbative matching across the thresholds corresponding to the charm- and bottom-quark masses, where perturbation theory works well, is $\alphasMSbar=0.11852(84)$ (for further technical details, see also refs.~\cite{DallaBrida:2016flw, DallaBrida:2016kgh} and the references therein). For a review of lattice results on $\alphas$, see ref.~\cite{Aoki:2019cca}.

Finally, a comparison of the results for $\alphas\left(m_{Z^0}\right)$ that we obtained in the theories with $N=2$ and $N=3$, when taken at face value, shows that $N \alphas\left(m_{Z^0}\right)$ is independent of $N$, to a sub-percent level of precision. This relation has a natural interpretation in terms of the 't~Hooft coupling in the large-$N$ limit of QCD~\cite{'tHooft:1973jz} (see also refs.~\cite{Allton:2008ty, Lucini:2008vi}), and it is unsurprising that it holds even for values of $N$ as small as $N=2$~\cite{Lucini:2012gg}. However, it is worth remarking that, even if one only compares purely gluonic theories, this statement may still depend on the way the physical scale is set.

\subsection{Computational efficiency analysis}
\label{subsec:computational_efficiency_analysis}

In this subsection, we assess the computational efficiency of our algorithm, and compare it with conventional (equilibrium) Monte~Carlo algorithms for the calculation of the running coupling in the Schr\"odinger-functional scheme. We first discuss general features and expectations, then we present quantitative details obtained from the analysis of our results.

The first important observation is that, at least superficially, our numerical determination of the running coupling follows quite closely the computation that is carried out in conventional simulations, in the sense that it is based on the same quantity, i.e. the derivative of the effective action $\Gamma$ with respect to $\eta$. In conventional algorithms, this derivative (which is nothing but a sum of chromoelectric field components evaluated on the Euclidean-time-slices $x_0=0$ and $x_0=L$: see, e.g., ref.~\cite[eq.~(3.2)]{Luscher:1992zx}) is evaluated directly, on an ensemble of equilibrium configurations. The calculation boils down to computing a sum of local objects (traces of plaquette operators with the insertion of a generator of the gauge-group algebra, as explicitly detailed in ref.~\cite[eq.~(3.3)]{Luscher:1992zx}) on the two boundaries of the system at the initial and final Euclidean times. Correspondingly, in our calculation, we estimated the effective-action derivative from the $\Delta \eta \to 0$ limit of the $\Delta \Gamma / \Delta \eta$ quotient ratio, extracting $\Delta \Gamma$ from eq.~(\ref{Jarzynski_theorem}), which requires the numerical evaluation of the variation in Euclidean action $\Delta S$ in each non-equilibrium trajectory. In turn, the variation in Euclidean action induced by a sequence of changes in $\eta$ is simply expressed in terms of the difference in the plaquette values at the boundaries.

Note, however, that, as we remarked in section~\ref{sec:numerical_implementation}, the total Euclidean-action variation during each trajectory is decomposed into the sum of $\nqq$ terms (each of them corresponding to the Euclidean-action variation induced by a single ``quantum quench''), hence one might be tempted to conclude that our non-equilibrium algorithm becomes exceedingly more time-consuming than a conventional Monte~Carlo simulation for large values of $\nqq$ (as discussed above, we used $\nqq=1000$ for most of our production runs). This argument, however, is misleading, because it does not take the \emph{distribution} of values of the observable into account.

To clarify this point, it is enlightening to discuss our non-equilibrium algorithm in two different, and particularly interesting, limits.

Firstly, we note that, for $\nqq=1$, $\eta$ would be immediately switched from its initial to its final value: then, according to eq.~(\ref{discretized_Euclidean-action_difference}), our algorithm would simply use a set of equilibrium configurations generated at $\eta=\eta(\tin)$ to compute the exponential average of the difference in Euclidean action obtained by changing the boundary fields from $\eta=\eta(\tin)$ to $\eta=\eta(\tfin)$. Then, eq.~(\ref{Jarzynski_theorem}) would reduce to
\begin{equation}
\label{reweighting}
\left\langle \exp \left\{ - \left[ S_{\eta(\tin)+ \Delta \eta}-S_{\eta(\tin)} \right] \right\} \right\rangle = \frac{Z_{\eta(\tin)+ \Delta \eta}}{Z_{\eta(\tin)}},
\end{equation}
with the average on the left-hand side evaluated on the equilibrium configurations generated at $\eta=\eta(\tin)$. This would then correspond to considering the exponential of the Euclidean-action difference between the two ensembles as an observable to be evaluated on the ensemble with $\eta=\eta(\tin)$, i.e. to an implementation of reweighting~\cite{Ferrenberg:1988yz}. Reweighting is a computational technique that is widely used in Monte~Carlo simulations of statistical-mechanics systems, whose derivation does not involve any non-equilibrium assumptions, and whose computational efficiency may be limited by the existence of a (possibly severe) overlap problem: when the typical configurations contributing to $Z_{\eta(\tin)+ \Delta \eta}$ and $Z_{\eta(\tin)}$ are very different, the evaluation of the left-hand side of eq.~(\ref{reweighting}) requires exceedingly large statistics. Note that one of the works we compare our results with, namely ref.~\cite{Luscher:1993gh}, pointed out the existence of a long tail in the distribution of the derivative of $\Gamma$ with respect to $\eta$ at the largest coupling considered there, leading to long auto-correlation times for this observable, and presented a detailed discussion of how this numerical problem was tackled using the reweighting technique.

Secondly, we note that in the opposite limit $\nqq \to \infty$, the $\eta$ parameter would be varied \emph{infinitely slowly}. As a consequence, the system would remain in equilibrium throughout the Monte~Carlo evolution: from a statistical-mechanics viewpoint, this would then imply that the work $W$ done on the system during each trajectory would be exactly equal to the difference between the free energies of the final and of the initial ensemble. In particular, this would also imply that the \emph{width} of the distribution of $W/T$ values contributing to the average on the left-hand side of eq.~(\ref{Jarzynski_theorem}) would tend to zero (so that, in principle, one could determine the $Z_{\lambda(\tfin)}/Z_{\lambda(\tin)}$ ratio from \emph{just one} trajectory).

We now proceed to a more quantitative study of these aspects, presenting a detailed analysis of our lattice results. We start from the simulation ensembles whose results are plotted in figure~\ref{fig:plot_of_distribution_Ncol_2_nt_6_nx_5_ny_5_nz_5_beta_2.7124000000_nsteps_200_eta_0.7853981634} and summarized in table~\ref{tab:su2_results}: as discussed in subsection~\ref{subsec:results_for_the_SU2_theory}, they are obtained from non-equilibrium simulations of $\SU(2)$ Yang-Mills theory on a lattice with $L/a=5$ at fixed $\beta=2.7124$ and for fixed $\nqq=200$, from $\eta(\tin)=\pi/4$ to $\eta(\tfin)=\eta(\tin)+\Delta \eta$, for the runs labeled as ``direct'' transformations (or vice~versa, for those denotes as ``reverse'' transformations), for different values of $\Delta \eta$. As mentioned in subsection~\ref{subsec:results_for_the_SU2_theory}, the distributions of Euclidean-action differences $\Delta S$ observed along the non-equilibrium trajectories spanned in those runs are approximated well by Gau{\ss}ian distributions. This is made manifest by the analysis of the first few moments and cumulants of these distributions (up to the fourth order): the mean, the variance, the standardized skewness and the excess kurtosis. For a real-valued variable $X$ with normalized probability distribution $\mathcal{P}(X)$, they are respectively defined as
\begin{eqnarray}
&& \mu_1 = \int X \mathcal{P}(X) \dd X,\\
&& s^2 = \int (X-\mu_1)^2 \mathcal{P}(X) \dd X, \\
&& \tilde{\mu}_3 = \frac{\int (X-\mu_1)^3 \mathcal{P}(X) \dd X}{s^3}, \\
&& \kappa-3 = \frac{\int (X-\mu_1)^4 \mathcal{P}(X) \dd X}{s^4}-3.
\end{eqnarray}
We estimated these measures for the Euclidean-action differences $\Delta S$ observed in our non-equilibrium simulations, obtaining the results reported in table~\ref{tab:moments_Ncol_2_nt_6_nx_5_ny_5_nz_5_beta_2.7124000000_nsteps_200_eta_0.7853981634}, where the values that are statistically compatible with zero are denoted by a star.

\begin{table}
\begin{tabular}{|c|l|c|l||c|l|c|l|}
\hline
 type & \multicolumn{1}{|c|}{$\Delta \eta$} & measure & \multicolumn{1}{|c||}{value} & type & \multicolumn{1}{|c|}{$\Delta \eta$} & measure & \multicolumn{1}{|c|}{value} \\
\hline \hline
direct & $0.015$ & $\mu_1$         & $-0.15680(4)$            & reverse & $0.015$ & $\mu_1$         & $0.15682(4)$ \\
       &         & $s^2$           & $0.000054(13)$           &         &         & $s^2$           & $0.000055(13)$ \\
       &         & $\tilde{\mu}_3$ & $0.02^\star$             &         &         & $\tilde{\mu}_3$ & $-0.03^\star$ \\
       &         & $\kappa-3$      & $0.03^\star$             &         &         & $\kappa-3$      & $0.05^\star$ \\
\hline
direct & $0.01$  & $\mu_1$         & $-0.104812(25)$          & reverse & $0.01$  & $\mu_1$         & $0.104873(27)$ \\
       &         & $s^2$           & $0.000025(5)$            &         &         & $s^2$           & $0.000024(6)$ \\
       &         & $\tilde{\mu}_3$ & $-0.004^\star$           &         &         & $\tilde{\mu}_3$ & $-0.002^\star$  \\
       &         & $\kappa-3$      & $0.04^\star$             &         &         & $\kappa-3$      & $-0.02^\star$  \\
\hline
direct & $0.005$ & $\mu_1$         & $-0.052544(12)$          & reverse & $0.005$ & $\mu_1$         & $0.052541(12)$ \\
       &         & $s^2$           & $0.0000061(13)$          &         &         & $s^2$           & $0000062(13)$ \\
       &         & $\tilde{\mu}_3$ & $0.012^\star$            &         &         & $\tilde{\mu}_3$ & $-0.02^\star$  \\
       &         & $\kappa-3$      & $0.016^\star$            &         &         & $\kappa-3$      & $0.04^\star$  \\
\hline
direct & $0.002$ & $\mu_1$         & $-0.021054(6)$           & reverse & $0.002$ & $\mu_1$         & $0.021056(6)$ \\
       &         & $s^2$           & $0.00000097(25)$         &         &         & $s^2$           & $00000099(27)$ \\
       &         & $\tilde{\mu}_3$ & $-0.006^\star$           &         &         & $\tilde{\mu}_3$ & $-0.003^\star$  \\
       &         & $\kappa-3$      & $0.07^\star$             &         &         & $\kappa-3$      & $0.09^\star$  \\
\hline
direct & $0.001$ & $\mu_1$         & $-0.010531(3)$           & reverse & $0.001$ & $\mu_1$         & $0.0105311(24)$ \\
       &         & $s^2$           & $00000025(6)$            &         &         & $s^2$           & $0.00000024(5)$ \\
       &         & $\tilde{\mu}_3$ & $0.02^\star$             &         &         & $\tilde{\mu}_3$ & $-0.01^\star$  \\
       &         & $\kappa-3$      & $0.04^\star$             &         &         & $\kappa-3$      & $-0.02^\star$  \\
\hline
\end{tabular}
\caption{Moments of the Euclidean-action difference $\Delta S$ obtained in non-equilibrium simulations of $\SU(2)$ Yang-Mills theory at fixed $L/a=5$,  $\beta=2.7124$ and $\nqq=200$, and for five different values of $\Delta \eta$, in direct and in reverse non-equilibrium transformations. Values denoted by a star are compatible with zero within their uncertainties.}
\label{tab:moments_Ncol_2_nt_6_nx_5_ny_5_nz_5_beta_2.7124000000_nsteps_200_eta_0.7853981634}
\end{table}

The table shows very clearly that, as expected, the width of the distribution of $\Delta S$ values obtained along non-equilibrium trajectories shrinks to zero when $\Delta \eta/\nqq$ tends to zero at fixed $\nqq$: as anticipated, this is the limit in which the field configurations spanned during the trajectories do not depart strongly from equilibrium, hence all values of the Euclidean-action difference measured numerically by our algorithm are very close to each other, and, according to eq.~(\ref{Jarzynski_theorem}), to the effective-action difference that is induced by that $\Delta \eta$. The fact that these trajectories are indeed quite close to equilibrium is also confirmed by another observation, namely that the Euclidean-action-difference distributions observed in ``reverse'' transformations (with statistical measures on the same lines of table~\ref{tab:moments_Ncol_2_nt_6_nx_5_ny_5_nz_5_beta_2.7124000000_nsteps_200_eta_0.7853981634}) are exactly symmetric, within the uncertainties on the parameters. Note that, in general, this is not always the case: for  simulations in which the system is driven to deviate strongly from equilibrium, the distribution of $\Delta S$ obtained in ``direct'' transformations \emph{is not} equal to the distribution of $-\Delta S$ measured in ``reverse'' transformations---on the contrary, the two distributions are expected to cross each other at one point $\Delta S=\Delta \Gamma$: in the setup that we are considering, this is nothing but the statement of Crooks' theorem~\cite{Crooks:1997ne}, from which eq.~(\ref{Jarzynski_theorem}) can be immediately derived.

Coming back to the discussion of the results listed in table~\ref{tab:moments_Ncol_2_nt_6_nx_5_ny_5_nz_5_beta_2.7124000000_nsteps_200_eta_0.7853981634}, we also note that all of the distributions analyzed there are characterized by values of the skewness and of the excess kurtosis compatible with zero, within their uncertainties. For example, for the data set reported in the first cell of the table, corresponding to ``direct'' transformations with $\Delta \eta=0.015$, we found $\tilde{\mu}_3 \simeq 0.02$, with an uncertainty (estimated by jackknife binning) of a few units, and $\kappa-3 \simeq 0.03$, with an even larger uncertainty, and similar results persist for the other data sets. On the one hand, the fact that both the skewness and the excess kurtosis vanish, confirms that in the (near-equilibrium) regime probed by those simulations, the Euclidean-action distributions measured along the trajectories are approximated very well by Gau{\ss}ians (as suggested by figure~\ref{fig:plot_of_distribution_Ncol_2_nt_6_nx_5_ny_5_nz_5_beta_2.7124000000_nsteps_200_eta_0.7853981634}). Note, also, that the large uncertainties on $\tilde{\mu}_3$ and on $\kappa-3$ are not due to limited statistics (the number of independent trajectories analyzed for each data set reported in table~\ref{tab:moments_Ncol_2_nt_6_nx_5_ny_5_nz_5_beta_2.7124000000_nsteps_200_eta_0.7853981634} is of the order of $4 \cdot 10^4$), but to the very limited amount of data that deviate significantly from normal distributions; it is also worth remembering that both $\tilde{\mu}_3$ and $\kappa-3$ are suitably normalized parameters, whose definitions encode fine cancellations among terms related to different moments of the distribution they characterize.

We conclude that the data in table~\ref{tab:moments_Ncol_2_nt_6_nx_5_ny_5_nz_5_beta_2.7124000000_nsteps_200_eta_0.7853981634} support the expectation that the distributions of values of $\Delta S$ can be modeled by normal distributions, whose mean values scale linearly with $\Delta \eta$ (see also figure~\ref{fig:SU2_Delta_Gamma_over_Delta_eta}). The width of these distributions tends to zero when the simulations approach the equilibrium limit, which happens when $\nqq$ is large for a given $\Delta \eta$: in that limit, the distributions of $\Delta S$ from direct transformations become symmetric with respect to those from reverse transformations, and tend to sharp distributions centered exactly at $\Delta \Gamma$.

For completeness, in table~\ref{tab:moments_Ncol_3_lower_left-hand-side_block} we report the results of the cumulant analysis for a set of simulations of the $\SU(3)$ gauge theory. In this case, the data were obtained from non-equilibrium simulations with $\Delta \eta=0.0001$ and $\nqq=1000$ fixed. As before, the results show clearly that the distributions of $\Delta S$ are very sharply peaked, that the results obtained from direct and reverse transformations are fully consistent with each other, and the parameters describing deviations from a Gau{\ss}ian shape are always consistent with zero.

\begin{table}
\begin{tabular}{|l|c||c|c|l||c|c|l|}
\hline
\multicolumn{1}{|c|}{$\beta$} & $L/a$ & type & measure & \multicolumn{1}{|c||}{value} & type & measure & \multicolumn{1}{|c|}{value} \\
\hline \hline
$8.1555$ & $5$ & direct & $\mu_1$         & $0.00262457(6)$     & reverse & $\mu_1$         & $-0.00262455(6)$ \\
         &     &        & $s^2$           & $0.0000000006(3)$   &         & $s^2$           & $0.00000000064(29)$ \\
         &     &        & $\tilde{\mu}_3$ & $-0.0008^\star$     &         & $\tilde{\mu}_3$ & $0.002^\star$ \\
         &     &        & $\kappa-3$      & $-0.007^\star$      &         & $\kappa-3$      & $0.01^\star$ \\
\hline
$8.3124$ & $6$ & direct & $\mu_1$         & $0.00263004(10)$    & reverse & $\mu_1$         & $-0.00262996(10)$ \\
         &     &        & $s^2$           & $0.0000000009(5)$   &         & $s^2$           & $0.0000000009(5)$ \\
         &     &        & $\tilde{\mu}_3$ & $0.01^\star$        &         & $\tilde{\mu}_3$ & $-0.002^\star$ \\
         &     &        & $\kappa-3$      & $-0.002^\star$      &         & $\kappa-3$      & $-0.007^\star$ \\
\hline
$8.4442$ & $7$ & direct & $\mu_1$         & $0.0026304(14)$     & reverse & $\mu_1$         & $-0.0026303(15)$ \\
         &     &        & $s^2$           & $0.000000002^\star$ &         & $s^2$           & $0.000000002^\star$ \\
         &     &        & $\tilde{\mu}_3$ & $-9.6^\star$        &         & $\tilde{\mu}_3$ & $12^\star$ \\
         &     &        & $\kappa-3$      & $250^\star$         &         & $\kappa-3$      & $430^\star$ \\
\hline
$8.5598$ & $8$ & direct & $\mu_1$         & $0.002628(6)$       & reverse & $\mu_1$         & $-0.002623(6)$ \\
         &     &        & $s^2$           & $0.000000007^\star$ &         & $s^2$           & $0.00000001^\star$ \\
         &     &        & $\tilde{\mu}_3$ & $-14^\star$         &         & $\tilde{\mu}_3$ & $12^\star$ \\
         &     &        & $\kappa-3$      & $290^\star$         &         & $\kappa-3$      & $220^\star$ \\
\hline
\end{tabular}
\caption{Same as in table~\ref{tab:moments_Ncol_2_nt_6_nx_5_ny_5_nz_5_beta_2.7124000000_nsteps_200_eta_0.7853981634}, but for our simulations of $\SU(3)$ Yang-Mills theory summarized in the lower left-hand-side block of table~\ref{tab:su3_deltaeta_0.0001}, with $\Delta \eta=0.0001$ and $\nqq=1000$.}
\label{tab:moments_Ncol_3_lower_left-hand-side_block}
\end{table}

\begin{figure}[!htbp]
\begin{center}
\includegraphics*[width=\textwidth]{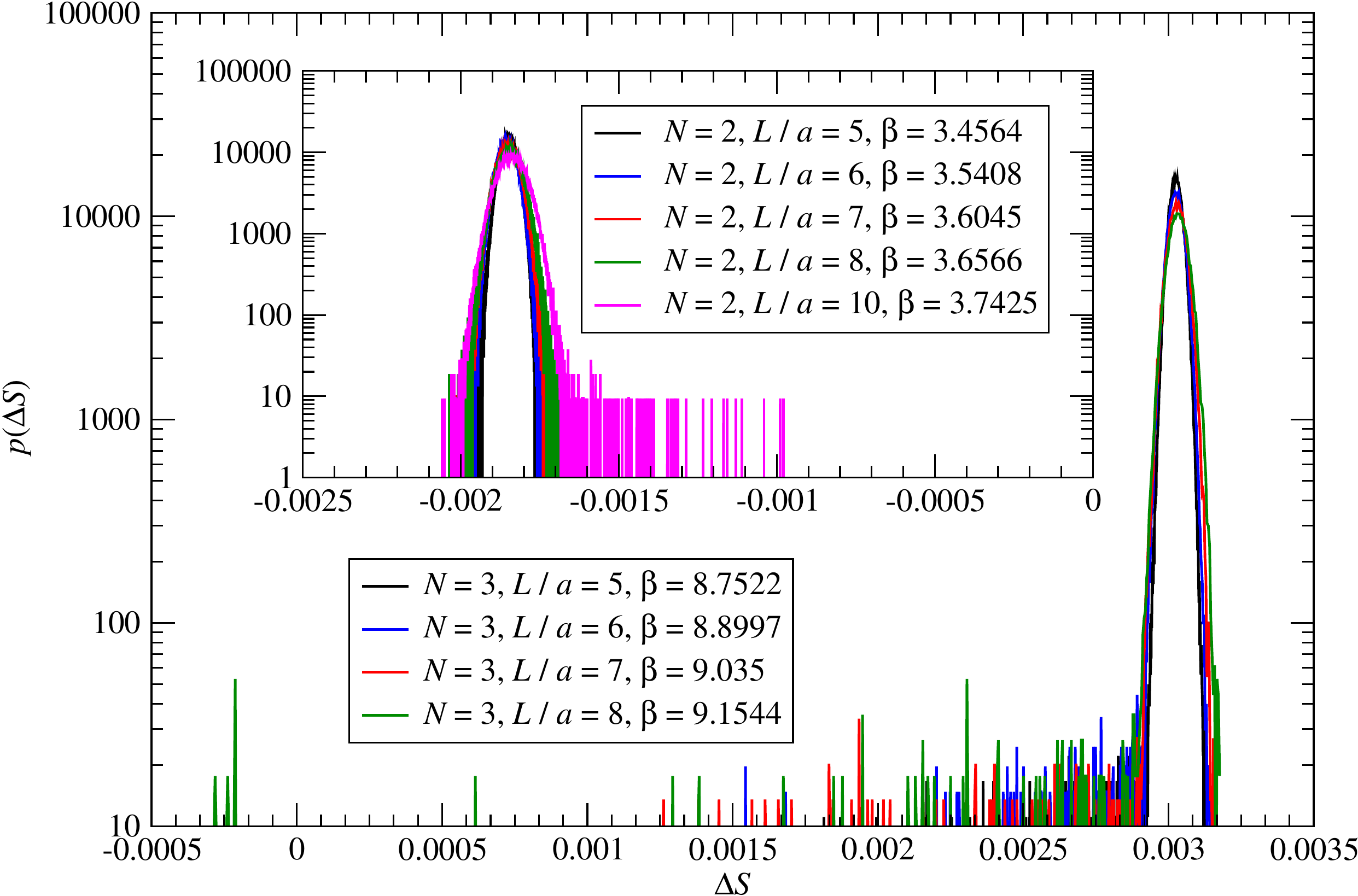}
\caption{\label{fig:distribution_comparison} The main plot shows a comparison of distributions for the Euclidean action difference $\Delta S$ obtained in ``direct'' non-equilibrium trajectories in $\SU(3)$ Yang-Mills theory, from a set of simulations on lattices with approximately constant $g^2(L)$ (i.e. approximately fixed physical size $L$), for different values of $L/a$, as listed in the first block of table~\ref{tab:su3_deltaeta_0.0001}. The inset shows an analogous sample of results for the $\SU(2)$ theory, obtained from the ``direct'' non-equilibrium trajectories summarized in the first block of table~\ref{tab:su2_deltaeta_0.0001}.}
\end{center}
\end{figure}

Finally, it is also interesting to note that, for a fixed physical size of the system, the shape of the normalized distribution of $\Delta S$ values depends only very mildly on the linear size of the system in units of the lattice spacing: this is shown in figure~\ref{fig:distribution_comparison}, where the main plot displays the results that we obtained for the $\SU(3)$ theory from the set of simulations summarized in the first block of table~\ref{tab:su3_deltaeta_0.0001}. The plot refers to the Euclidean-action difference obtained in direct non-equilibrium transformations, on lattices corresponding to $g^2(L) \simeq 1.247$, i.e. for a fixed physical size $L$, with $\Delta \eta=0.0001$, $\nqq=1000$, and for $L/a$ ranging from $5$ to $8$. The four distributions show a remarkable collapse to a common curve (up to very small deviations in regions far from the peak, which however are not significant within the statistical uncertainties, and a corresponding slight increase of $s^2$ with $L/a$), despite a nearly sevenfold increase in the number of degrees of freedom ($N^2-1$ for each of the link matrices not fixed by the Dirichlet boundary conditions discussed in section~\ref{sec:numerical_implementation}). Similar conclusions are obtained for the $\SU(2)$ theory: in this case, the inset in figure~\ref{fig:distribution_comparison} shows the distributions obtained from direct non-equilibrium transformations, again with $\Delta \eta=0.0001$ and $\nqq=1000$, on lattices of approximately fixed physical size $L$ at which $g^2(L) \simeq 2.04$, for $L/a$ values from $5$ to $10$ (corresponding to an increase in the number of degrees of freedom by a factor larger than $17$), that are listed in the first block of table~\ref{tab:su2_deltaeta_0.0001}. Also in this case, the distributions obtained numerically, from our non-equilibrium simulations, exhibit a remarkable collapse to the same curve (up to slight deviations in regions far from the maximum, where the observed distributions are, however, smaller by more than three orders of magnitude with respect to the region close to the peak, and not significant within the uncertainties). In addition, we note that the differences among the various distributions in the maximum region are perhaps slightly more visible than in the $\SU(3)$ case. This, however, is likely to be at least partially due to the slightly larger relative difference in the values of $g^2(L)$ for this set of $\SU(2)$ simulations: comparing the results shown in the first block of table~\ref{tab:su2_deltaeta_0.0001} and those in the first block of table~\ref{tab:su3_deltaeta_0.0001}, we observe that the relative variations in the $g^2(L)$ values are always well below $1\%$, but are slightly larger for the $\SU(2)$ theory than for the $\SU(3)$ theory.

These features can be compared and contrasted with those characterizing the distribution of values that are obtained by directly computing $\Delta \Gamma / \Delta \eta$ in ordinary, equilibrium Monte~Carlo simulations. In that case, there is no departure from equilibrium at all, and hence no parameter analogous to our $\nqq$; accordingly, the distribution of values of $\Delta \Gamma / \Delta \eta$ has its own, fixed, width and shape, which has to be efficiently sampled by the simulation algorithm. In many cases, this task can be carried out without any challenges by ordinary Monte~Carlo algorithms. For the $\SU(2)$ theory, in 1992 the authors of ref.~\cite{Luscher:1992zx} were able to achieve the significant level of precision of the results reported in that work using an amount of CPU time on CRAY~YMP processors ranging from approximately $35$~hours for the lattices of linear size in units of the lattice spacing $L/a \le 14$, to about $140$~hours for the lattices with $L/a=20$. For comparison, for the same theory the amount of CPU time that in this work we used to generate $3347$ trajectories on a lattice with $L/a=20$ and $\beta=3.7425$ on a single core of CINECA machines equipped with Intel Xeon Phi 7250 CPU (Knights Landing) processors at $1.40$~GHz is approximately $100$~days. While this comparison may appear very humbling for our non-equilibrium algorithm, it is important to remark that these trajectories were produced with a large $\nqq$ value ($\nqq=1000$), which, as can be appreciated from our final results, leads to enhanced precision for the value of the coupling that we could extract from them.

The efficiency of the non-equilibrium algorithm over standard equilibrium Monte~Carlo, however, becomes particularly manifest in cases when the latter has to sample long tails in the distribution of $\Delta \Gamma / \Delta \eta$, which lead to large autocorrelation times (as discussed, for example, in ref.~\cite{Luscher:1993gh} for simulations of the $\SU(3)$ theory at $g^2\simeq 3.48$). In a conventional equilibrium Monte~Carlo algorithm, this problem should necessarily be addressed by applying reweighting methods, or some other similar technique. It is under these circumstances that our non-equilibrium algorithm proves particularly competitive in terms of CPU costs, since it allows one to bypass the computational overhead that is typically associated with reweighting and other analogous techniques. More precisely, the fact that our algorithm drives the field configurations to evolve out of equilibrium allows them to efficiently probe regions of the phase space of the system that would be exponentially difficult to access by standard reweighting methods.

In other words, the non-equilibrium algorithm discussed in the present work can be considered as one that simultaneously generalizes both standard Monte~Carlo simulations and reweighting algorithms (recovering them in two particular limits, as discussed above). For quantities that can be efficiently estimated by conventional simulation algorithms, our code fares no better than them: although its design relies on a different approach, the physical observable it computes is, in fact, very similar to the one that is directly accessed by standard algorithms used for the determination of the Schr\"odinger-functional coupling, and the drawback of discretizing the trajectories into a sufficiently large number of steps is offset by the smaller fluctuations affecting the Euclidean-action differences measured along them. Under conditions where the dynamics of the theory is such, that conventional Monte~Carlo algorithms are hampered by long autocorrelation times, difficulties in sampling the configuration space, or ensemble-overlap problems, however, non-equilibrium simulations can provide a very efficient alternative.

Finally, it is worth remarking that the conclusions we discussed above should hold regardless of the precise functional form of the time-dependent variation protocol for the parameters of the system that are modified during the non-equilibrium transformations ($\eta$ in this case). While we restricted our analysis only to one class of such protocols, i.e. to sequences of ``quantum quenches'' of the same amplitude in Monte~Carlo time, it is easy to guess that alternative choices for $\eta(t)$ can have a significant impact on the computational efficiency of this type of non-equilibrium simulations. Interesting examples include a protocol $\eta_1(t)$ that remains constant to $\eta(\tin)$ until the last step, when $\eta$ is suddenly driven to $\eta(\tfin)$, and another, $\eta_2(t)$, in which, conversely, $\eta$ is immediately switched to $\eta(\tfin)$ at the beginning of the trajectory, but then remains constant for the rest of the trajectory. The former would not drive the field configurations out of equilibrium at all during the whole trajectory except at the last step, and it would then determine $\Delta \Gamma$ simply by reweighting the last ``measured'' configuration to $\eta=\eta(\tfin)$. By contrast, a protocol like $\eta_2(t)$ would closely mimic one sudden, large, initial ``quench'', which drives the field configurations violently out of equilibrium at the beginning of each trajectory, and then allows them to (tend to) relax towards the equilibrium state corresponding to $\eta=\eta(\tfin)$. More interesting examples could include, for instance, protocols $\eta(t)$ in which $\eta$ is neither constant nor linear: while their investigation is beyond the scope of the present work, they may be optimized to improve the computational efficiency of the non-equilibrium algorithm, particularly under dynamical conditions in which it is competitive with respect to conventional equilibrium Monte~Carlo simulations.

\section{Conclusions}
\label{sec:conclusions}

In this work, we presented the results of a non-perturbative study of the running coupling of non-Abelian gauge theories, by means of a non-equilibrium Monte~Carlo algorithm that implements a numerical realization of Jarzynski's theorem. Specifically, we evaluated the response in effective action induced by a deformation of the boundary conditions at the initial and final Euclidean time, and extracted the running coupling in the Schr\"odinger-functional scheme~\cite{Luscher:1992an}.

The latter scheme provides a well-defined formulation of the theory in a finite system, with Dirichlet boundary conditions along Euclidean time and periodic (or periodic up to a constant phase, for fermionic fields) boundary conditions along the three spatial directions, and allows one to define the renormalized coupling at a momentum scale defined as the inverse of the linear extent of the system in each direction, $\mu=1/L$. This formulation is amenable to lattice regularization and, through the iterative procedure that we discussed in section~\ref{sec:results_and_analysis}, it allows one to study the evolution of a renormalized quantity when the momentum scale varies by orders of magnitude, by recursively matching the (continuum-extrapolated) value of the coupling obtained on lattices of the same physical extent, but different lattice spacing. The application of this technique to study the non-perturbative renormalization in lattice QCD was pioneered in ref.~\cite{Jansen:1995ck}, where the renormalization of the axial current, the running coupling in the chiral limit, and the momentum-scale evolution of the renormalized axial density were discussed, as well as the non-perturbative determination of the coefficients of improvement terms to reduce lattice artifacts. Since then, the formalism has been successfully applied to study the renormalized gauge coupling and a number of other physical quantities in QCD~\cite{Sint:1998iq, Guagnelli:1998ve, Kurth:2000ki, Guagnelli:2005zc, DellaMorte:2005kg, Aoki:2005et, Palombi:2007dr, Dimopoulos:2007ht, Pena:2017hct} and, more recently, has also been used to investigate the dynamics of other strongly coupled non-supersymmetric gauge theories with dynamical fermionic fields~\cite{Appelquist:2007hu, Shamir:2008pb, DelDebbio:2008zf, Hietanen:2009az, Bursa:2009we, Hayakawa:2010yn, Karavirta:2011zg, Hayakawa:2013yfa, Fodor:2015zna}.

The goal of this work consisted in showing that the Schr\"odinger-functional formalism can be directly implemented in Monte~Carlo calculations \emph{out of equilibrium}, using the powerful fluctuation theorems that have been recently developed in statistical mechanics. As discussed in section~\ref{sec:numerical_implementation}, the central idea is to drive the system out of equilibrium through a sequence of ``quantum quenches in Monte~Carlo time'': Jarzynski's theorem~\cite{Jarzynski:1996oqb, Jarzynski:1997ef} implies that the \emph{exponential average} of the Euclidean-action variation induced in this process is directly related to the exponential of the difference in effective action between the initial and the final states of the system.

We emphasize that, while in our computation we evaluate a quantity (the discretized derivative of the effective action with respect to $\eta$, the parameter that specifies the Dirichlet boundary conditions of the system at the initial and final Euclidean times) which is directly related, and can be made arbitrarily close, to the one that is evaluated in conventional simulations of the Schr\"odinger functional, the approach is intrinsically radically different, as our calculation does not rest on the standard formalism of equilibrium Monte~Carlo calculations. In particular, our approach closely ``mimics'' the non-trivial dynamics induced in physical statistical systems that are experimentally driven out of equilibrium, and the corresponding measurements that can be performed on them. Experimental applications of this type are diverse, and include, for example, irreversible mechanical stretching of ribonucleic acid molecules~\cite{Liphardt:2002ei}.

We focused on $\SU(2)$ and $\SU(3)$ Yang-Mills theories, and showed that the results obtained in our non-equilibrium Monte~Carlo simulations are fully compatible with those from standard (equilibrium) lattice simulations~\cite{Luscher:1992zx, Luscher:1993gh}. While we presented results for purely bosonic theories, the generalization of this calculation to include dynamical fermions poses no additional conceptual challenge, and could be easily carried out with the same techniques common to lattice QCD, e.g. through (a non-equilibrium version of) the hybrid Monte~Carlo algorithm~\cite{Callaway:1982eb, Callaway:1983ee, Polonyi:1983tm, Duane:1985hz, Duane:1986iw}.

Finally, we mention some other recent, and very interesting, articles that present applications of non-equilibrium statistical-mechanics theorems in a context relevant for quantum field theory~\cite{Alba:2016bcp, Bartolotta:2017rth, Ortega:2019etm, DEmidio:2019usm}; in particular, refs.~\cite{Alba:2016bcp, DEmidio:2019usm} focus on the calculation of the entanglement entropy from Jarzynski's equality, whereas refs.~\cite{Ortega:2019etm, DEmidio:2019usm} discuss the implications of non-equilibrium theorems for quantum field theories. We expect that many more such studies, at the interface between modern statistical mechanics and quantum field theory, will appear in the near future, and that they may lead to new insights into open problems both in condensed matter theory and in elementary particle theory.

\vskip1.0cm 
\noindent{\bf Acknowledgements}\\
The work of O.F. is partially supported by the ANR Project No.~ANR-15-IDEX-02. The numerical simulations were run on machines of the Consorzio Interuniversitario per il Calcolo Automatico dell'Italia Nord Orientale (CINECA).

\bibliography{paper}

\end{document}